\documentclass[preprint,nofootinbib,aps,superscriptaddress,eqsecnum]{revtex4-1}
 \pdfoutput=1
\usepackage{amsmath,amssymb,graphicx,subfigure}

\newcommand{\beq}{\begin{equation}}
\newcommand{\eeq}{\end{equation}}
\newcommand{\bea}{\begin{eqnarray}}
\newcommand{\eea}{\end{eqnarray}}
\newcommand{\Z}{\mathbb{Z}}

\def\nn{\nonumber}

\def\ra{\rightarrow}
\def\mpl{M_{\rm Pl}}

\newcommand{\MS}{\overline{\mbox{\sc ms}}}

\def\etc{ {\em etc.\ }}
\def\ie{ {\em i.e.,\ }}

\def\mpl{M_{\rm Pl}}

\begin{document}
% \begin{center}
\title{Constraints on inert dark matter from metastability of the electroweak vacuum }
%\vspace {1.0cm}

\author{Najimuddin Khan}\email{phd11125102@iiti.ac.in} 
\author{Subhendu Rakshit}\email{rakshit@iiti.ac.in} 
\affiliation{Discipline of Physics, Indian Institute of Technology Indore,   \\
IET-DAVV Campus, Indore 452017, India \vspace{1.80cm}}

%\end{center}

\begin{abstract}\vspace*{10pt}
The inert scalar doublet model of dark matter can be valid up to the Planck scale. 
We briefly review the bounds on the model in such a scenario and identify  parameter spaces that lead to absolute stability and metastability  of the electroweak vacuum.
\end{abstract}

\renewcommand*{\thefootnote}{\fnsymbol{footnote}}
\maketitle
%\newpage

\section{Introduction}
Although the LHC has unearthed the standard model (SM) Higgs boson~\cite{Aad:2012tfa,Chatrchyan:2012ufa,Giardino:2013bma}, so far, it has failed to identify any sign of new physics beyond the SM.  New physics at the TeV scale is invoked in the form of supersymmetry, extra dimensions \etc to address the issue of the hierarchy problem. Absence of such new physics can imply that the solution of the hierarchy problem lies somewhere else. Hence, it is likely that the SM could be valid up to the Planck scale ($\mpl$), above which gravity is expected to dominate over other forces.

However, we know that the presence of dark matter (DM) and nonzero neutrino masses indicate physics beyond the SM. Hence, it is worthwhile to explore a scenario where the SM, augmented by a low energy DM model, is valid up to $\mpl$. Earlier we had  explored such a scenario where a gauge singlet scalar is added to the SM to pose as a weakly interacting massive particle (WIMP) DM~\cite{Khan:2014kba}. In the present work, we extend the SM by a scalar doublet protected by a $\Z_2$ symmetry. The model is popularly known in the literature as the inert doublet~(ID) model, first proposed by Deshpande and Ma~\cite{Deshpande:1977rw}. 

Given the experimental measured values for $M_t$, $M_h$, and $\alpha_s$, the electroweak (EW) vacuum is reported to reside in a metastable state within the framework of the SM. This has been verified in the literature~\cite{Bezrukov:2012sa, Degrassi:2012ry, Masina:2012tz, Buttazzo:2013uya, Spencer-Smith:2014woa} only very recently using state of the art next-to-next-to leading order (NNLO) loop corrections contributing to the Higgs effective potential. In Ref.~\cite{Khan:2014kba}, this analysis was extended to the case of a singlet scalar DM model. The stability of the EW vacuum was shown to depend on new physics parameters. 
In this paper, we extend such an analysis to the ID model.  We assume that ID DM is the only DM particle which saturates the entire DM relic density. In this context, we review the constraints on the parameters of the ID model. 

A detailed study on the ID parameter space was recently performed in Refs.~\cite{Dolle:2009fn, Gustafsson:2012aj} indicating bounds from EW stability, perturbativity, collider study, electroweak precision tests (EWPT), etc., considering DM annihilation to two-body final states only. In Refs.~\cite{Honorez:2010re, Gustafsson:2012aj} DM relic density was calculated including three-body final states as well. An updated detailed analysis on the ID parameter space was performed in Refs.~\cite{Goudelis:2013uca, Arhrib:2013ela} reflecting the impact of the $M_h\approx 126$~GeV Higgs boson discovery at the LHC. The authors of Refs.~\cite{Gustafsson:2012aj, Goudelis:2013uca} looked at renormalization group evolution (RGE) of model parameters to check the validity of the ID model at higher energies.  The constraints from the measurements of the diphoton decay channel of the Higgs boson on the ID parameter space was also discussed.  High scale validity of this model in the presence of a right-handed neutrino has been looked at in Ref.~\cite{Chakrabarty:2015yia}. The influence of ID in the Higgs effective potential was studied in Ref.~\cite{Hambye:2007vf} to explore the possibility of  electroweak symmetry breaking {\it \`a la} Coleman-Weinberg~\cite{Coleman:1973jx}.  High scale validity of the two Higgs doublet model has been explored in Refs.~\cite{Chakrabarty:2014aya, Das:2015mwa} with broken $\Z_2$, symmetry considering tree level potential only. 

In this paper we improve earlier studies on ID model parameter space by including radiative corrections to the scalar potential to explore the stability of the electroweak vacuum. In particular, we permit the Higgs quartic coupling to assume small negative values to render a metastable EW vacuum. As our analysis improves the scalar potential considering radiative corrections, the tree level stability constraints used in earlier analyses need to be reviewed. In addition, as we demand that the theory be valid up to $\mpl$, the model parameters at the EW scale are constrained by the requirement that they satisfy all the bounds up to the $\mpl$. If in the near future the direct detection experiments and colliders confirm the ID model as the dark matter model realized in nature, then our study will help to estimate the lifetime of the electroweak vacuum, especially if it still remains in the metastable state. As metastability of EW vacuum is considered to be an indication of new physics, if the vacuum is metastable even after consideration of the ID model, then that could indicate the presence of additional new physics before $\mpl$.

The paper is organized as follows. Section~\ref{sec:IDM} starts with an introduction of the ID model, followed by a discussion of the effective scalar potential and renormalization group (RG) running of parameters of the model. Constraints on the model are discussed in Sec.~\ref{sec:constraints}. Tunneling probability of the metastable vacuum and the restrictions on the parameters to avoid a potential unbounded from below, are mentioned in Sec.~\ref{sec:metastability}. A detailed study of the parameter space identifying regions of EW vacuum stability and metastability is performed in Sec.~\ref{sec:phasediag} with the help of various phase diagrams. Section~\ref{sec:veltman} contains a short discussion on Veltman's conditions in the context of the ID model. We finally conclude in Sec.~\ref{sec:conclusion}.

\section{Inert Doublet Model}
\label{sec:IDM}
In this model, the standard model is extended by adding an extra $SU(2)$ doublet scalar, odd under an additional discrete $\Z_2$ symmetry. Under this symmetry, all standard model fields are even. The $\Z_2$ symmetry prohibits the inert doublet to acquire a vacuum expectation value.

The scalar potential at the tree level is given by 
\bea
V(\Phi_1,\Phi_2) &=& \mu_1^2 |\Phi_1|^2 + \lambda_1 |\Phi_1|^4+ \mu_2^2 |\Phi_2|^2 + \lambda_2 |\Phi_2|^4\nn\\
&&+ \lambda_3 |\Phi_1|^2 |\Phi_2|^2 
+  \lambda_4 |\Phi_1^\dagger \Phi_2|^2 + \frac{\lambda_5}{2} \left[ (\Phi_1^\dagger \Phi_2)^2 + {\rm H.c.}\right]  \, ,
\label{Scalarpot}
\eea
where the SM Higgs doublet $\Phi_1$ and the inert doublet $\Phi_2$ are given by
\beq
	\Phi_1 ~=~ \left( \begin{array}{c} G^+ \\ \frac{1}{\sqrt{2}}\left(v+h+i G^0\right) \end{array} \right),
	\qquad
	\Phi_2 ~=~ \left( \begin{array}{c} H^+\\ \frac{1}{\sqrt{2}}\left(H+i A\right) \end{array} \right) \, \nn.
\eeq
$G^\pm$ and $G^0$ are Goldstone bosons and $h$ is the SM Higgs. 

$\Phi_2$ contains a $CP$ even neutral scalar $H$, a $CP$ odd neutral scalar $A$,  and a pair of charged scalar fields $H^\pm$. The $\Z_2$ symmetry prohibits these particles to decay entirely to SM particles. The lightest of $H$ and $A$ can then serve as a DM candidate. 

After electroweak symmetry breaking, the scalar potential is given by
\bea
V(h, H,A,H^\pm) &=&  \frac{1}{4} \left[ 2 \mu_1^2 (h+v)^2 + \lambda_1 (h+v)^4 +2 \mu_2^2 (A^2+H^2+2 H^+ H^-) \right. \nn \\
&& \left. + \lambda_2 (A^2 + H^2 + 2 H^+ H^-)^2  \right] \nn \\
&& + \frac{1}{2} (h+v)^2 \left[  \lambda_3 H^+ H^- 
+  \lambda_S  A^2  
+  \lambda_L  H^2 \right] \label{Scalarpot2}
\eea
where 
\bea
\lambda_{L,S}&=&\frac{1}{2}\left(\lambda_3+\lambda_4\pm\lambda_5\right) \, .
\eea
Masses of these scalars are given by
\begin{align}
	M_{h}^2 &= \mu_1^2 + 3 \lambda_1 v^2,\nn \\
	M_{H}^2 &= \mu_2^2 +  \lambda_L v^2, \nn \\
	M_{A}^2 &= \mu_2^2 + \lambda_S v^2,\nn \\
	M_{H^\pm}^2 &= \mu_2^2 + \frac{1}{2} \lambda_3 v^2  \,\nn .
\end{align}

For  $\lambda_4-\lambda_5<0$ and $\lambda_5>0$~($\lambda_4+\lambda_5<0$ and $\lambda_5<0$), %implies that 
$A$~($H$) is the lightest $\Z_2$ odd particle (LOP). In this work, we take $A$ as the LOP and hence, as a viable DM candidate. Choice of $H$ as LOP will lead to similar results. 

As we will explain later, for large DM mass, $M_A\gg M_Z$, appropriate relic density of DM is obtained if $M_A$, $M_H$, and $M_{H^\pm}$ are nearly degenerate. Hence, in anticipation, we define
\bea
	\Delta M_H&=& M_H-M_A, \nn\\
	\Delta M_{H^\pm}&=&M_{H^\pm}-M_A \nn\, .
\eea
so that the new independent parameters for the ID model become $\{M_A, \Delta M_H, \Delta M_{H^\pm}, \lambda_2, \lambda_S\}$. Here we have chosen $\lambda_S$ as  we treat $A$ as the DM particle.

The one-loop effective potential for $h$ in the $\MS$ scheme and the Landau gauge is given by   
\beq
V_1^{{\rm SM}+{\rm ID}}(h)= V_1^{\rm SM}(h) + V_1^{\rm ID}(h)
\eeq
where
\beq
V_1^{\rm SM}(h)=\sum_{i=1}^5 \frac{n_i}{64 \pi^2} M_i^4(h) \left[ \ln\frac{M_i^2(h)}{\mu^2(t)}-c_i\right] \, .
\eeq
$n_i$ is the number of degrees of freedom. For scalars and gauge bosons, $n_i$ is positive, whereas for fermions it is  negative. Here $c_{h,G,f}=3/2$, $c_{W,Z}=5/6$, and the running energy scale $\mu$ is expressed in terms of a dimensionless parameter $t$ as $\mu(t)=M_Z \exp(t)$. $M_i$ is given as 
\beq
M_i^2(h)= \kappa_i(t)\, h^2(t)-\kappa_i^{\prime}(t) \, .
\nn \eeq
$n_i$, $\kappa_i$ and $\kappa_i^{\prime}$ can be found in Eq.~(4) in Ref.~\cite{Casas:1994qy} (see also  Refs.~\cite{Altarelli:1994rb, Casas:1994us, Casas:1996aq, Quiros:1997vk}).

The additional contribution to the one-loop effective potential due to the inert doublet is given by~\cite{Hambye:2007vf}
\beq
V_1^{\rm ID}(h)= \sum_{j=H,A,H^+,H^-} \frac{1}{64 \pi^2} M_j^4(h) \left[ \ln\left(\frac{M_j^2(h)}{\mu^2(t)} \right)- \frac{3}{2} \right] 
\eeq
where 
\beq
M_j^2(h)=\frac{1}{2} \,\lambda_{j}(t) \, h^2(t)+\mu_2^2(t) 
\eeq
with $\lambda_{A}(t)=2 \lambda_{S}(t)$, $\lambda_{H}(t)=2 \lambda_{L}(t)$,   and $\lambda_{H^\pm}(t)= \lambda_{3}(t)$.

In the present work, in the Higgs effective potential, SM contributions are taken at the two-loop level~\cite{Degrassi:2012ry, Buttazzo:2013uya,Ford:1992pn,Martin:2001vx}, whereas the ID scalar contributions are considered at one loop only. 

For $h \gg v$, the Higgs effective potential can be approximated as
\beq
V_{\rm eff}^{{\rm SM}+{\rm ID}}(h) \simeq \lambda_{\rm 1,eff}(h) \frac{h^4}{4}\, ,
\label{efflam}\eeq
with
\beq
\lambda_{\rm 1,eff}(h) = \lambda_{\rm 1,eff}^{\rm SM}(h) +\lambda_{\rm 1,eff}^{\rm ID}(h)\, ,
\eeq
where~\cite{Buttazzo:2013uya}
\bea
\lambda_{\rm 1,eff}^{\rm SM}(h) &=& e^{4\Gamma(h)}
\left[ \lambda_1(\mu=h) + \lambda_{\rm 1,eff}^{(1)}(\mu=h) +  \lambda_{\rm 1,eff}^{(2)}(\mu=h)\right] \nn\\
 \lambda_{\rm 1,eff}^{\rm ID}(h)&=&\sum_{j=L,S,3} e^{4\Gamma(h)} \left[\frac{\delta_j \lambda_j^2}{64 \pi^2}  \left(\ln\left(\delta_j\lambda_j\right)-\frac{3}{2}\right ) \right]\, . \nn
\eea
Here $\delta_j=1$ when $j=L,S$;  $\delta_j=\frac{1}{2}$ for $j=3$; and
\beq
\Gamma(h)=\int_{M_t}^{h} \gamma(\mu)\,d\,\ln\mu \, .\nn
%\label{Gamma}\\
\eeq
Anomalous dimension $\gamma(\mu)$ of the Higgs field takes care of  its wave function renormalization.
As quartic scalar interactions do not contribute to wave function renormalization at the one-loop level, ID does not alter  $\gamma(\mu)$ of the SM.  The expressions for the one- and two-loop quantum corrections $\lambda_{\rm 1,eff}^{(1,2)}$ in the SM can be found in Ref.~\cite{Buttazzo:2013uya}. All running coupling constants are evaluated at $\mu=h$. This choice guarantees that the perturbative expansion of the potential is more reliable~\cite{Ford:1992mv}, which can be understood as follows. The perturbative expansion of the potential at high field values of $h$ consists of terms of the form $\epsilon\,\ln(h/\mu)$, where $\epsilon$ is some dimensionless coupling. Clearly the perturbation series works better if we choose the renormalization scale $\mu \sim h$ and ensure that $\epsilon$ is small as well. Hence, our choice of $\mu=h$, along with the requirement that all the couplings remain within the perturbative domain, ensures faster convergence of the perturbation series of the potential.

To compute the RG evolution of all the couplings, we first calculate all the SM couplings at $M_t$, taking care of the threshold corrections as in Refs.~\cite{Sirlin:1985ux,Bezrukov:2012sa,Degrassi:2012ry,Khan:2014kba}. Then we evolve them up to $\mpl$ using our own computer codes incorporating the RG equations~\cite{Chetyrkin:2012rz,Zoller:2012cv,Chetyrkin:2013wya,Zoller:2013mra}. The corresponding $\beta$-functions include three-loop SM effects and two-loop ID contributions (one-loop RGEs are presented in the Appendix for completeness).  If DM mass $M_A$ is larger than $M_t$, then the ID starts to contribute after the energy scale $M_A$. For  $M_A<M_t$, the contributions of ID to the $\beta$-functions are rather negligible for the running from $M_A$ to $M_t$, as is evident from the expressions. 

\begin{figure}[h!]
 \begin{center}
 % \subfigure[]{
 \includegraphics[width=3.7in,height=3.1in, angle=0]{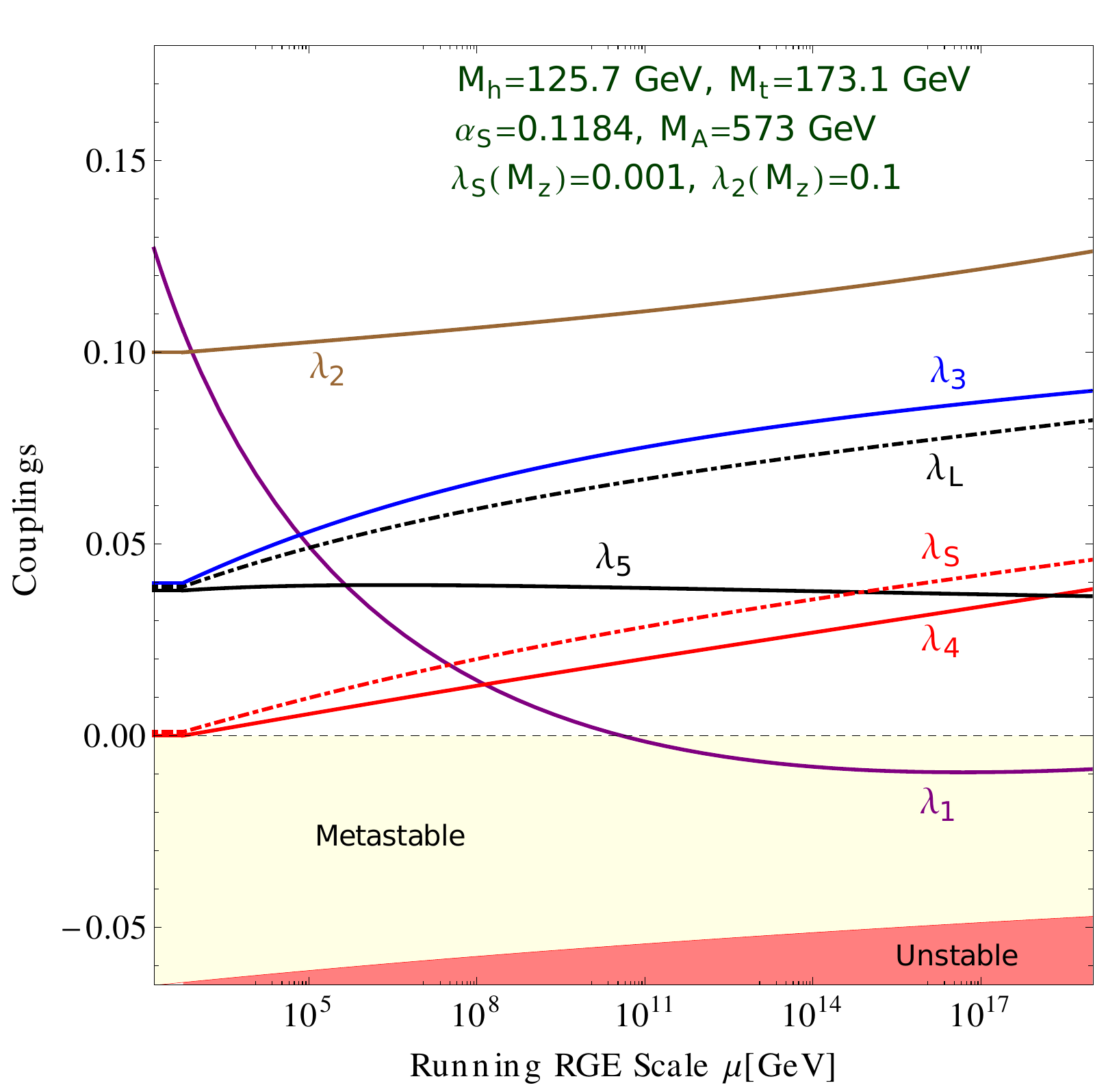}
 \caption{\label{fig:SMInert} \text{RG evolution of the couplings $\lambda_i\left(i=1,..,5\right), \lambda_L, \lambda_S$ for the set of parameters in Table~\ref{table1}. } }
 \end{center}
 \end{figure}

To help the reader in reproducing our results, we provide in Table~\ref{table1}  a specific set of values of $\lambda_i$ at $M_t=173.1$~GeV and at $\mpl = 1.2 \times 10^{19}$~GeV for $M_h=125.7$~GeV and $\alpha_s\left(M_Z\right)=0.1184$.  In Fig.~\ref{fig:SMInert} we explicitly show running of the scalar couplings $(\lambda_i)$ for this set of parameters. We see that for this specific choice of parameters, $\lambda_1$ assumes a small negative value\footnote{$\lambda_1$ and $\lambda_{\rm 1,eff}$, being almost equal at $\mpl$, are interchangeable at that scale.} leading to a metastable EW vacuum as discussed in the following sections. This set is chosen to reproduce the DM relic density in the right ballpark.
%%%%%%%%%
\begin{table}[h!]
\begin{center}
    \begin{tabular}{| c | c | c | c | c | c | c | c |}
    \hline
     &$\lambda_S$ & $\lambda_L$ & $\lambda_2$ & $\lambda_3$ & $\lambda_4$ & $\lambda_5$ & $\lambda_1$\\
\hline
   ~$M_{t}$ ~&~ 0.001 ~&~ 0.039 ~&~ 0.10 ~&~ 0.0399 ~&~ 0.00003   ~&~ 0.038 ~&~ 0.127~\\
            \hline
   $\mpl$ & 0.046 & 0.082 & 0.127 & 0.090 & 0.038   & 0.036 & $-0.009$\\
              \hline
    \end{tabular}
    \caption{A set of values of all ID model coupling constants at  $M_t$ and $\mpl$ for $M_{A}=573$ GeV, $\Delta M_{H^\pm}=1$ GeV, $\Delta M_{H}=2$ GeV, and $\lambda_S\left(M_Z\right)=0.001$.}
    \label{table1}
\end{center}
\end{table}
%%%%%%%%%

%%%%RGE RUNNING
%%%%RGE RUNNING

\newpage

\section{Constraints on ID model: A brief review}
\label{sec:constraints}

ID model parameter space is constrained from theoretical considerations like  absolute vacuum stability, perturbativity and unitarity of the scattering matrix.  Electroweak precision measurements and direct search limits at LEP put severe restrictions on the model. The recent measurements of Higgs decay width at the LHC put additional constraints. The requirement that the ID DM saturates the DM relic density all alone restricts the allowed parameter space considerably. Although these bounds are already discussed in the literature~\cite{LopezHonorez:2006gr}, here we apply these bounds with the requirement that the model needs to be valid till $\mpl$.

\subsection{Vacuum stability bounds}

The tree level scalar potential potential $V(\Phi_1,\Phi_2)$ is stable and bounded from below if~\cite{Deshpande:1977rw} 
\beq
\lambda_{1,2}(\Lambda) \geq 0, \quad  \lambda_{3}(\Lambda) \geq -2 \sqrt{ \lambda_{1}(\Lambda)\lambda_{2}(\Lambda)}, \quad  \lambda_{L,S}(\Lambda) \geq - \sqrt{ \lambda_{1}(\Lambda)\lambda_{2}(\Lambda)}  \label{stabilitybound}
\eeq
where the coupling constants are evaluated at a scale $\Lambda$ using RG equations. However, these conditions become nonfunctional if $\lambda_1$ becomes negative at some energy scale to render the EW vacuum metastable. Under such circumstances we need to handle metastability constraints on the potential differently, which we pursue in the next section. 

\subsection{Perturbativity bounds}

The radiatively improved scalar potential remains perturbative by requiring that all quartic couplings of $V(\Phi_1,\Phi_2)$ satisfy 
\beq
\mid \lambda_{1,2,3,4,5}\mid \leq 4 \pi %;\lbrace i = 1,2,3,4,5\rbrace
\eeq

\subsection{Unitarity bounds}

Unitarity bounds on  $\lambda_i$ are obtained considering scalar-scalar, gauge boson-gauge boson, and scalar-gauge boson scatterings~\cite{Lee:1977eg}. The constraints come from the eigenvalues of the corresponding S-matrix~\cite{Arhrib:2012ia}:
\bea
\vert \lambda_3 \pm \lambda_4\vert\leq 8\pi , ~~~~~\vert\lambda_3 \pm \lambda_5\vert\leq 8\pi\nn\\
\vert \lambda_3+ 2 \lambda_4 \pm 3\lambda_5\vert\leq 8 \pi\nn\\
\vert-\lambda_1 - \lambda_2 \pm \sqrt{(\lambda_1 - \lambda_2)^2 + \lambda_4^2}\vert\leq 8 \pi
\\
\vert-3\lambda_1 - 3\lambda_2 \pm \sqrt{9(\lambda_1 - \lambda_2)^2 + (2\lambda_3 + \lambda_4)^2}\vert\leq 8 \pi
\nn\\ 
\vert -\lambda_1 - \lambda_2 \pm \sqrt{(\lambda_1 - \lambda_2)^2 + \lambda_5^2}\vert\leq 8 \pi\nn
\label{unitary}
\eea

\subsection{Bounds from electroweak precision experiments}

Bounds ensuing from electroweak precision experiments are imposed on new physics models {\it via} Peskin-Takeuchi~\cite{Peskin:1991sw} $S, T, U$ parameters. The additional contributions from ID are given by~\cite{Barbieri:2006dq,Arhrib:2012ia}
\beq
 \Delta S = \frac{1}{2\pi}\Bigg[ \frac{1}{6}\ln\left(\frac{M^2_{H}}{M^2_{H^\pm}}\right) -
  \frac{5}{36} + \frac{M^2_{H} M^2_{A}}{3(M^2_{A}-M^2_{H})^2} + 
\frac{M^4_{A} (M^2_{A}-3M^2_{H})}{6(M^2_{A}-M^2_{H})^3} \ln \left(\frac{M^2_{A}}{M^2_{H}}\right)\Bigg]
\label{Sparam}
\eeq
and
\bea
 \Delta T &=& \frac{1}{32\pi^2 \alpha v^2}\Bigg[ F\left(M^2_{H^\pm}, M^2_{A}\right)
+ F\left(M^2_{H^\pm}, M^2_{H}\right) - F\left(M^2_{A}, M^2_{H}\right)\Bigg]
\label{Tparam}
\eea
where the function $F$ is given by
\beq
	F(x,y) = \left\{ \begin{array}{lr}
		\frac{x+y}{2} - \frac{xy}{x-y}\ln\left(\frac{x}{y}\right), & x\neq y \\
		0, & \, x=y
		\end{array}  \right. \, .
\eeq
We use the NNLO global electroweak fit results obtained by the Gfitter group~\cite{Baak:2014ora}, 
\beq
\Delta S = 0.06 \pm 0.09, ~ ~ \Delta T = 0.1 \pm 0.07
\label{STU1}
\eeq
with a correlation coefficient of $+0.91$, fixing $\Delta U$ to zero.  We use this fit result as the contribution of the scalars in the ID model to $\Delta U$ is rather negligible.

%%%%ContourSTU
 \begin{figure}[h!]
 \begin{center}
% \subfigure[]{
 \includegraphics[width=2.8in,height=2.8in, angle=0]{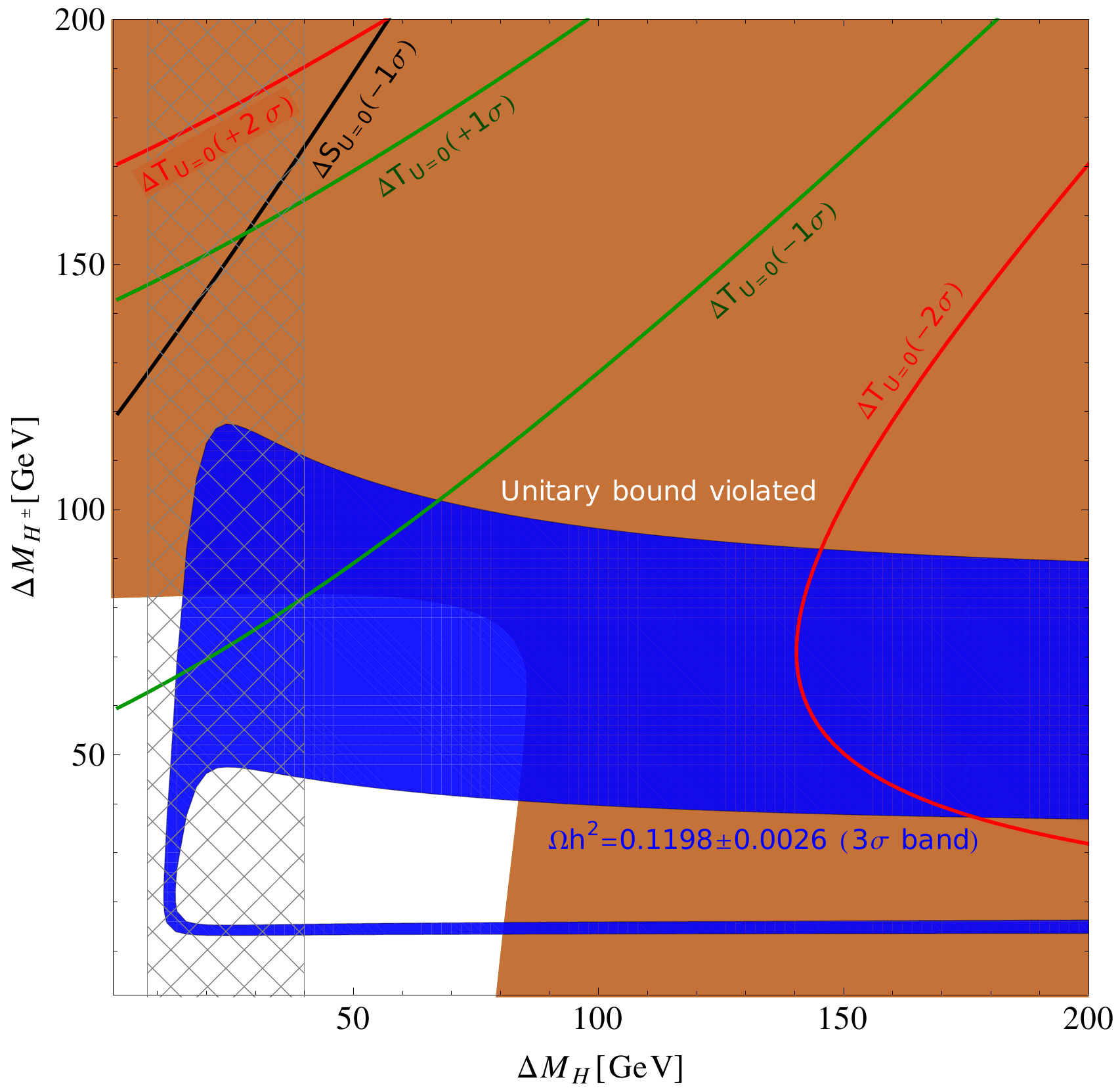}
  \caption{\label{fig:STUcheck} \textit{\rm{Allowed parameter space in $\Delta M_{H^\pm}-\Delta M_{H}$ plane for $M_A=70$~GeV and $\lambda_S=0.007$. Constraints from $S$ and $T$ parameters are shown by solid black, green and red lines. The blue band corresponds the 3$\sigma$ variation in $\Omega h^2=0.1198\pm0.0026$~\cite{Ade:2013zuv}. On the brown region the unitarity bound is violated on or before $\mpl$. The cross-hatched region is excluded from LEP\,II data. }} } 
 \end{center}
 \end{figure} 
%%%%ContourSTU
To assess the implications of $S$ and $T$ constraints on the ID model, in Fig.~\ref{fig:STUcheck}, in the $\Delta M_{H^\pm}-\Delta M_{H}$ plane, we display various constraints for $M_A=70$~GeV and $\lambda_S=0.007$. This is the maximum value of $\lambda_S$ for the given DM mass, allowed by LUX~\cite{Akerib:2013tjd} direct detection data at 1$\sigma$.  The blue region allowed by relic density constraints shifts upwards for smaller $\lambda_S$. Constraints on $\Delta S$ and $\Delta T$, as mentioned in Eq.~(\ref{STU1}),  are marked as black, green, and red  solid lines. We see that the $1\sigma$ bound on $\Delta T$ is the most stringent one. The black line corresponding to the lower limit on $\Delta S$ at $1\sigma$ can also be interesting. But the LEP\,II bounds, represented by the cross-hatched bar, take away a considerable part.  The line representing the $\Delta S$ upper limit is beyond the region considered and lies towards the bottom-right corner of the plot.  In this plot, increasing $M_H$ enhances $S$, but $T$ gets reduced. In addition, if we demand unitarity constraints to be respected up to $\mpl$, assuming no other new physics shows up in between, the parameter space gets severely restricted. In this plot, a small window is allowed by $\Delta T$ only at $2\sigma$, which satisfies DM relic density constraints. As mentioned earlier, this window gets further reduced with smaller $\lambda_S$ (the relic density band takes an ``L'' shape as shown in Fig.~\ref{fig:STU2}). 

%However, such severe electroweak precision constraints are artefacts of  using Eqn.~(\ref{STU1}). If we use Eqn.~(\ref{STU2}) instead, some parameter space is allowed by  $\Delta T$ even at $1\sigma$, as indicated by the dashed purple lines. 

\subsection{Direct search limits from LEP}

The decays $Z\ra AH$, $Z\ra H^+ H^-$, $W^\pm\ra A H^\pm$, and $W^\pm\ra H H^\pm$ are restricted from $Z$ and $W^\pm$ decay widths at LEP. It implies $M_{A} + M_{H} \geq M_{Z}$, $2M_{H^\pm}\geq M_{Z}$, and $M_{H^\pm} + M_{H,A} \geq M_{W}$. More constraints on the ID model can be extracted from chargino~\cite{Pierce:2007ut} and neutralino~\cite{Lundstrom:2008ai} search results at LEP\,II: The charged Higgs mass $M_{H^\pm}\geq 70$~GeV. The bound on $M_A$ is rather involved: If  $M_A<80$ GeV, then $\Delta M_H$ should be less than $\sim 8$ GeV, or else $M_H$ should be greater than $\sim 110$~GeV ({see Fig.~\ref{fig:STU1}}).

\subsection{Bounds from LHC}
In the ID model, Higgs to diphoton signal strength $\mu_{\gamma\gamma}$ is defined as 
\beq
\mu_{\gamma\gamma} = \frac{\sigma(gg\ra h\ra\gamma\gamma)}{\sigma(gg\ra h\ra\gamma\gamma)_{\rm SM}}\approx \frac{Br(h \rightarrow {\gamma\gamma})_{\rm ID}}{Br(h \rightarrow {\gamma\gamma})_{\rm SM}}
\eeq
using the narrow width approximation for the production cross section of $\sigma(gg\ra h\ra \gamma\gamma)$ and the fact that $\sigma(gg\rightarrow h)$ in both the SM and ID are the same. 

Now if the ID particles have masses less than $M_h/2$, $h\rightarrow \rm ID,ID$ decays are allowed. In that case,
\beq 
 \mu_{\gamma\gamma}= \frac{\Gamma(h\rightarrow \gamma\gamma)_{\rm ID}} {\Gamma(h\rightarrow \gamma\gamma)_{\rm SM}} ~ \frac{\Gamma_{\rm tot}(h\rightarrow \rm SM,SM)}{\Gamma_{\rm tot}(h\rightarrow \rm SM,SM)+\Gamma_{\rm tot}(h\rightarrow \rm ID,ID)}\, ,
 \label{mugagalow}
\eeq
where~\cite{Cao:2007rm} 
\beq
\Gamma\left(h\rightarrow \rm ID,ID\right)= \frac{ v^2}{16\pi M_h }\lambda_{\rm ID}^2 \left(1-\frac{4 M_{\rm ID}^2}{M_{h}^2}\right)^{1/2}\, ,
\label{decaywidth}
\eeq
where for ${\rm ID}=A,H,H^\pm$, $\lambda_{\rm ID}=\lambda_S, \lambda_L, \sqrt{2}\,\lambda_3$.

In this case, the ID particles are heavier than $M_h/2$, 
\beq 
 \mu_{\gamma\gamma}= \frac{\Gamma(h\rightarrow \gamma\gamma)_{\rm ID}}{\Gamma(h\rightarrow \gamma\gamma)_{\rm SM}}\, .
 \label{mugagahigh}
\eeq
In the ID model, the $H^\pm$ gives additional contributions at one loop. 
The analytical expression is given by~\cite{Djouadi:2005gj, Swiezewska:2012eh,Krawczyk:2013jta}
\beq
\Gamma(h\rightarrow \gamma\gamma)_{\rm ID}={\alpha^2 m_h^3\over 256\pi^3
v^2}\left|\sum_{f}N^c_fQ_f^2y_f
F_{1/2}(\tau_f)+ y_WF_1(\tau_W)
+Q_{H^{\pm}}^2{v\mu_{{hH^+H^-}}\over
2m_{H^{\pm}}^2}F_0(\tau_{H^{\pm}})\right|^2\
\label{hgaga}
\eeq
where $\tau_i=m_h^2/4m_i^2$. $Q_{f}$, $Q_{H^{\pm}}$ denote electric charges of corresponding particles. $N_f^c$ is the color factor. $y_f$ and
$y_W$ denote Higgs couplings to $f\bar{f}$ and $WW$. $\mu_{hH^+H^-}= \lambda_3 v$ stands for the coupling constant of the $hH^+H^-$ vertex. The loop functions $F_{(0,\,1/2,\,1)}$  are defined as
\begin{align}
F_{0}(\tau)&=-[\tau-f(\tau)]\tau^{-2}\, ,\nn\\
F_{1/2}(\tau)&=2[\tau+(\tau-1)f(\tau)]\tau^{-2}\, ,\nn\\
F_{1}(\tau)&=-[2\tau^2+3\tau+3(2\tau-1)f(\tau)]\tau^{-2}\, , \nn
\label{loopfn}
\end{align}
where 
\beq
f(\tau)= \bigg\{\begin{array}{ll}
(\sin^{-1}\sqrt{\tau})^2\,,\hspace{60pt}& \tau\leq 1\\
-{1\over4}[\ln{1+\sqrt{1-\tau^{-1}}\over1-\sqrt{1-\tau^{-1}}}-i\pi]^2\,,\quad
&\tau>1
\label{ftau}
\end{array}\;,\;
\eeq

From the diphoton decay channel of the Higgs at the LHC, the measured values are  $\mu_{\gamma\gamma}$=$1.17\pm0.27$ from ATLAS~\cite{Aad:2014eha} and  $\mu_{\gamma\gamma}$=$1.14^{+0.26}_{-0.23}$ from CMS~\cite{Khachatryan:2014ira}. 

One can see that a positive $\lambda_3$ leads to a destructive interference between SM and ID contributions in En.~(\ref{hgaga}) and {\it vice versa}. Hence, for ID particles heavier than $M_h/2$, $\mu_{\gamma\gamma}<1$ ($\mu_{\gamma\gamma}>1$) when $\lambda_3$ is positive (negative). However, if these ID particles happen to be lighter than $M_h/2$, they might contribute to the invisible decay of the Higgs boson. Using the global fit result~\cite{Belanger:2013xza} that such an  invisible branching ratio is less than $\sim 20$\%, in Eq.~(\ref{mugagalow}), the second ratio provides a suppression of $\sim 0.8$ -- $1$.

Now can we work with a negative $\lambda_3$ in the ID model? We will discuss this at the end of this section. For the benchmark points used in this paper, we have worked only with positive values of $\lambda_3$, allowed at 1$\sigma$ by both CMS and ATLAS experiments. It is important to note that the constraints from $\mu_{\gamma\gamma}$ pass through the sign of $\lambda_3$ in our vacuum stability considerations.

\subsection{Constraints from dark matter relic density and direct search limits}

The ID dark matter candidate $A$ can self-annihilate into SM fermions. Once the DM mass is greater than the $W$-mass, so that the DM can annihilate into a pair of $W$ bosons, the cross section increases significantly, thereby reducing DM relic density. Hence, it becomes difficult to saturate $\Omega h^2$ after $\sim 75$~GeV with positive $\lambda_S$, although for $M_A<75$~GeV, both signs of  $\lambda_S$ can be allowed to arrive at the right DM relic density $\Omega h^2$.

The role of the sign of $\lambda_S$ can be understood from the contributing diagrams to the  $AA\ra W^+ W^-$  annihilation processes. Four diagrams contribute: the $AAW^+ W^-$ vertex driven point interaction diagram (henceforth referred to as the $p$-channel diagram),  $H^+$-mediated $t$- and $u$-channel diagrams, and the $h$-mediated $s$-channel diagram. For $AA\ra ZZ$  annihilation, the $t$- and $u$-channel diagrams are mediated by $H$. A negative $\lambda_S$ induces a destructive interference between the $s$-channel diagram with the rest, thereby suppressing $AA\ra W^+ W^-, ZZ$ processes. For DM masses of $75-100$~GeV, this can be used to get the appropriate $\Omega h^2$~\cite{Gustafsson:2012aj, LopezHonorez:2010tb}. To avoid large contributions from $t$- and $u$-channel diagrams and coannihilation diagrams, the $M_H$ and $M_{H^\pm}$ can be pushed to be rather large $\gtrsim 500$~GeV. However, to partially compensate the remaining $p$-channel diagram by the  $s$-channel one, $\lambda_S$ assumes a large negative value $\sim -0.1$, which is ruled out by the DM direct detection experiments. That is why in the ID model, DM can be realized below 75 GeV, a regime we designate as the ``low'' DM mass region.

At ``high'' DM mass $M_A \gtrsim 500$~GeV, one can get the right $\Omega h^2$ due to a partial cancellation between different diagrams contributing to the $AA\ra W^+ W^-$ and $AA\ra ZZ$ annihilation processes. For example, in $AA\ra W^+ W^-$, the $p$-channel diagram tends to cancel with the $H^+$-mediated $t$- and $u$-channel diagrams~\cite{Hambye:2009pw, Goudelis:2013uca} in the limit $M_A\gg M_W$, and the sum of amplitudes of these diagrams in this limit is proportional to $M_{H^\pm}^2-M_A^2$. Hence, at high $M_A$, a partial cancellation between these diagrams is expected for nearly degenerate $M_{H^\pm}$ and $M_A$. Similarly, for $AA\ra ZZ$, a cancellation is possible when the masses $M_{H}$ and $M_A$ are close by. For $M_A \gtrsim 500$~GeV, keeping the mass differences of $M_{H^\pm}$ and $M_{H}$ with $M_A$ within $8$~GeV, such  cancellations help reproduce the correct $\Omega h^2$. It is nevertheless worth mentioning that such nearly degenerate masses will lead to coannihilation of  these $\Z_2$ odd ID scalars~\cite{griest} to SM particles. Despite such near degeneracy, both $H$ and $H^+$, being charged under the same $\Z_2$ as $A$, decay promptly to the LOP $A$, so that they do not become relics.  We use {\tt FeynRules}~\cite{Alloul:2013bka} along with {\tt micrOMEGAs}~\cite{Belanger:2010gh, Belanger:2013oya} to compute the relic density of $A$. 

DM direct detection experiments involve the $h$-mediated $t$-channel process $AN\ra AN$ with a cross section proportional to $\lambda_S^2/M_A^2$ in the limit $M_A\gg M_N$: 
\beq
 \sigma_{A,N} = \frac{m_r^2}{\pi} f^2 m_N^2\left(\frac{\lambda_S}{M_{A} M_h^2}\right)^2 \label{directcs}
\eeq
where $f\approx0.3$ is the form factor of the nucleus. $m_r$ represents the reduced mass of the nucleus and the scattered dark matter particle. 

Thus, $\lambda_S$ is constrained from nonobservation of DM signals at XENON\,100~\cite{Aprile:2011hi,Aprile:2012nq} and LUX~\cite{Akerib:2013tjd}. For $M_{A}= 70$~GeV, the ensuing bound from LUX~\cite{Akerib:2013tjd,dmtools} data at 1$\sigma$ is $|\lambda_S|< 0.007$.  

The constraint on $\lambda_S$ from DM direct detection experiments gets diluted with $M_A$ [see Eq.~(\ref{directcs})]. Hence, for low DM mass, direct detection bounds are more effective. At high mass, the relic density constraints are likely to supersede these bounds. For example, for $M_{DM}= 573$~GeV, the upper limit on $|\lambda_S|$ is 0.138 from LUX. However, to satisfy the relic density constraints from the combined data of WMAP and Planck within 3$\sigma$, $\lambda_S$ can be as large as $0.07$ only.

Within the framework of the ID model, it is possible to explain the observations in various indirect DM detection experiments~\cite{Arhrib:2013ela, Modak:2015uda} for some regions of the parameter space. In this paper, however, we do not delve into such details as such estimations involve proper understanding of the astrophysical backgrounds and an assumption of the DM halo profile which contain some arbitrariness. For a review of constraints on the ID model from astrophysical considerations see, for example, Ref.~\cite{Gustafsson:2010zz}.

\vskip 30pt
\noindent{\bf\underline {Sign of $\lambda_3$}}\\

Whether $\lambda_3$ can be taken as positive or negative depends on the following:
\begin{itemize}

\item
If the ID model is not the answer to the DM puzzle, so that both relic density and direct detection constraints can be evaded, no restriction exists on the possible sign of $\lambda_3(M_Z)$. Otherwise, the following two cases need be considered:
\begin{itemize}
\item
A negative $\lambda_3(M_Z)$ implies 
\beq
\lambda_S(M_Z) < -\frac{1}{v^2} (M_{H^\pm}^2-M_A^2) \nn\, .
\eeq
As we are considering $A$ as the DM candidate, so that $M_A<M_{H^\pm}$, $\lambda_S(M_Z)$ is always negative when $\lambda_3(M_Z)<0$ . 
For low DM mass, the splitting  $(M_{H^\pm}-M_A)\gtrsim 10$~GeV, as otherwise DM coannihilation processes cause an inappreciable  depletion in $\Omega h^2$. For $M_A=70$~GeV, this implies a lower bound $\lambda_S(M_Z)\lesssim -0.025$, which violates the DM direct detection bound $|\lambda_S|< 0.007$. Hence, for low DM mass, a negative $\lambda_3(M_Z)$  is not feasible.
\item
For high mass DM, the right relic density can be obtained when the splitting  $(M_{H^\pm}-M_A)\sim$ a few GeV or less. The above logic then implies that a negative $\lambda_3$ does not put any severe restriction on $\lambda_S$ to contradict DM direct detection bounds as earlier. Hence, for a high DM mass, $\lambda_3(M_Z)$ can assume both the signs. Moreover, due to propagator suppression for large $M_{H^\pm}$ in the $h\gamma\gamma$ vertex, the ID contribution to $\mu_{\gamma\gamma}$ is negligibly small and hence, the sign of $\lambda_3(M_Z)$ is not constrained by measurements on $\mu_{\gamma\gamma}$ as well. 
\end{itemize}
\item
If at any scale, $\lambda_3$ is negative while $\lambda_1>0$, then the bound~(\ref{stabilitybound}) must be respected. 
\item
If at some scale, $\lambda_1<0$, then a negative $\lambda_3$ makes the potential unbounded from below, as mentioned in the following section. This means one can start with a negative $\lambda_3(M_Z)$, but with RG evolution when $\lambda_1$ turns negative, $\lambda_3$ evaluated at that scale must be positive. Such parameter space does exist. Here, we note a significant deviation of our analysis from earlier analyses which did not allow a negative $\lambda_1$. For example, in Ref.~\cite{Goudelis:2013uca} a negative $\lambda_3(M_Z)$ was not allowed from stability of the Higgs potential if the theory has to be valid up to $10^{16}$~GeV together with relic density considerations. 
\end{itemize}

\section{Tunneling Probability and Metastability}
\label{sec:metastability}
 %%%%tunneling probability
 \begin{figure}[h!]
 \begin{center}
\subfigure[]{
 \includegraphics[width=2.8in,height=2.8in, angle=0]{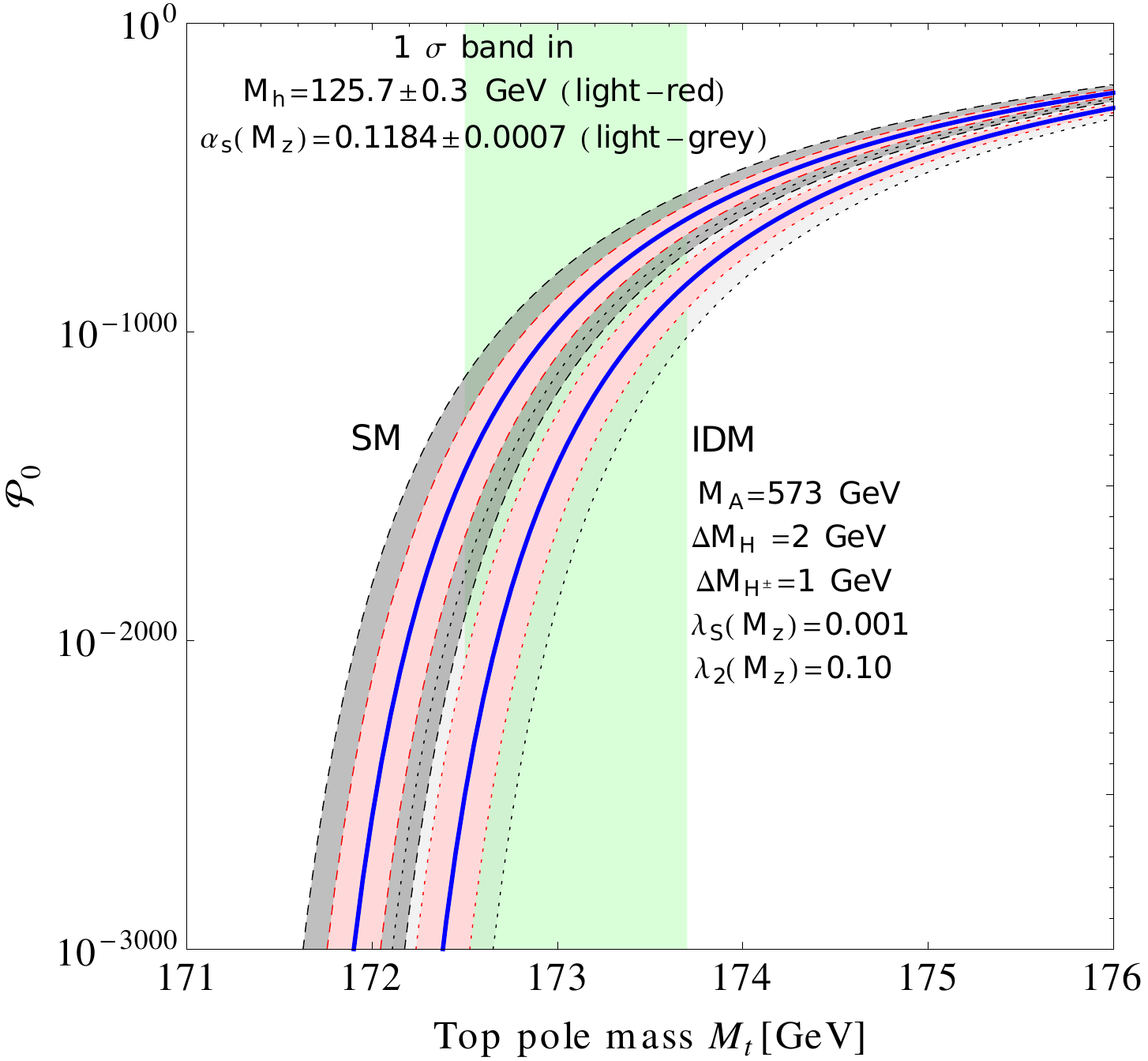}}
 \hskip 15pt
 \subfigure[]{
 \includegraphics[width=2.8in,height=2.8in, angle=0]{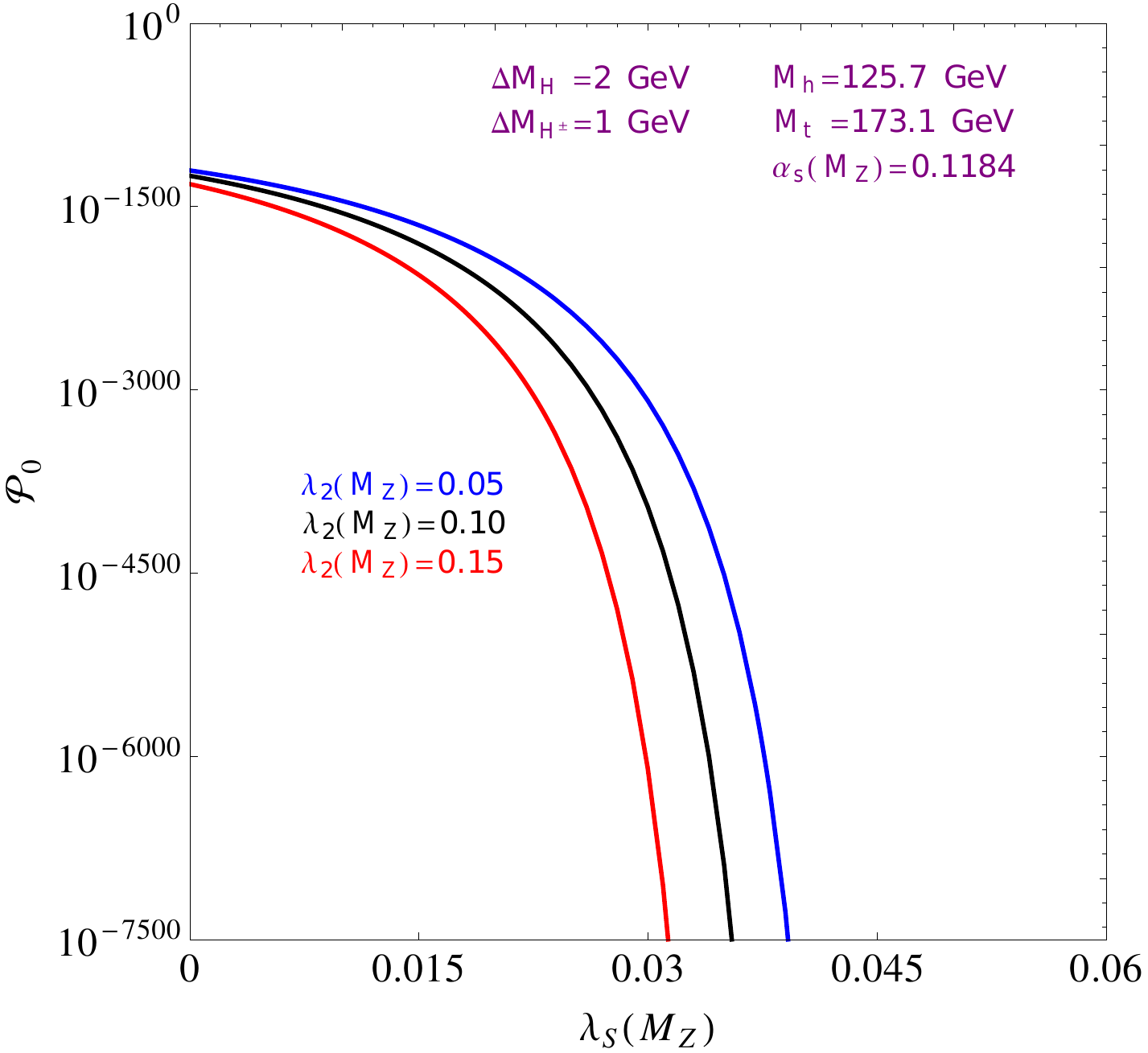}}
 \caption{\label{fig:Tun} \textit{ \rm{(a) Tunneling probability ${\cal P}_0$ dependence on  $M_t$. The left band (between dashed lines) corresponds to {\rm SM}. The right one (between dotted lines) is for $ID$ model for DM mass $M_A=573$~GeV.  Constraints from WMAP and Planck measured relic density, as well as XENON\,100 and LUX DM direct detection null results are respected for these specific choice of parameters. Light-green band stands for $M_t$ at $\pm 1\sigma$. (b) ${\cal P}_0$ is plotted against Higgs dark matter coupling $\lambda_S(M_Z)$ for different values of $\lambda_2(M_Z)$.}}}
 \end{center}
 \end{figure}
%%%%tunneling probability
In the standard model, the present measurements on $M_h$ and $M_t$ indicate that the electroweak vacuum, in which the Universe is at present residing, may be a false one. From this metastable vacuum, it might tunnel into a deeper true vacuum, residing close to $\mpl$. 

The decay probability of the EW vacuum to the true vacuum at the present epoch can be expressed as~\cite{Coleman:1977py, Isidori:2001bm, Buttazzo:2013uya}
\beq
{\cal P}_0=0.15 \frac{\Lambda_B^4}{H^4} e^{-S(\Lambda_B)}
\label{prob}\\
\eeq
where $H$ is the Hubble constant and the action is given by
\beq
S(\Lambda_B)=\frac{8\pi^2}{3|\lambda_1(\Lambda_B)|}\, .
\label{action}\\
 \eeq
$S(\Lambda_B)$ is called the action of bounce of size $R=\Lambda_B^{-1}$. The value of $R$ for which  $S(\Lambda_B)$  is minimum gives the dominant contribution to the tunneling probability ${\cal P}_0$. It occurs when $\lambda_1(\Lambda_B)$ is minimum, \ie $\beta_{\lambda_1}(\Lambda_B)=0$. Henceforth, $\Lambda_B$ denotes the scale where $\lambda_1$ is minimum. Here we neglect the loop correction to the action, as in Ref.~\cite{Isidori:2001bm} it had been argued that setting the running scale to $R^{-1}$ significantly restricts the size of such corrections. In the numerical evaluation we have used $\lambda_{1,\rm{eff}}$ in place of $\lambda_{1}$ in Eq.~(\ref{action}). We also neglect gravitational corrections~\cite{Coleman:1980aw,Isidori:2007vm} to the action as in Ref.~\cite{Khan:2014kba}. 
In Ref.~\cite{Isidori:2001bm} it was pointed out that thermal corrections are important at very high temperatures. 
Finite temperature effects to EW vacuum stability in the context of SM have been calculated recently in a preprint~\cite{Rose:2015lna}\footnote  {See Ref.~\cite{Espinosa:1995se} for an earlier work.}. It had been claimed that the parameter space corresponding to EW metastability shrinks considerably. In this paper we work with zero temperature field theory only. 

The presence of the inert doublet induces additional contributions to $\beta_{\lambda_1}$ [see Eq.~(\ref{betal_1})].  As a result, which is generic for all scalars,  $\lambda_1$ receives a positive contribution compared to the SM, which pushes a metastable vacuum towards stability, implying a lower ${\cal P}_0$. 

Electroweak metastability in the ID model has been explored earlier in the literature, albeit in a different context~\cite{Barroso:2013awa,Barroso:2012mj,Barroso:2013kqa,Gil:2012ya}. If $H^+$ gets a VEV, there could exist another charge-violating minimum. If instead, $A$ receives a VEV, another $CP$-violating minimum could pop up. But these vacua always lie higher than the usual EW vacuum. If $\Z_2$ is broken by introducing additional soft terms in the Lagrangian, then the new $\Z_2$-violating minimum can be lower than the usual $\Z_2$-preserving EW minimum. As in our present work, $\Z_2$ is an exact symmetry of the scalar potential, so such cases need not be considered. However, as mentioned earlier, if at some scale before $\mpl$, the sign of $\lambda_1$ becomes negative, there might exist a deeper minimum which is charge-, $CP$- and $\Z_2$-preserving and lying in the SM Higgs $h$ direction. 

Whether the EW vacuum is metastable or unstable, depends on the minimum value of $\lambda_1$ before $\mpl$, which can be understood as follows. For EW vacuum metastability, the decay lifetime should be greater than the lifetime of the Universe, implying ${\cal P}_0 < 1$. This implies~\cite{Isidori:2001bm, Khan:2014kba}
\beq
\lambda_{\rm 1, eff}(\Lambda_B) > \lambda_{\rm 1,min}(\Lambda_B)=\frac{-0.06488}{1-0.00986 \ln\left( {v}/{\Lambda_B} \right)}\,.
\label{lammin}
\eeq
Hence we can now reframe the vacuum stability constraints on the ID model, when $\lambda_{\rm 1, eff}$ runs into negative values, implying metastability of the EW vacuum. We remind the reader that in the ID model, instability of the EW vacuum cannot be realized as addition of the scalars only improves the stability of the vacuum. 
\begin{itemize}
\item
If $0>\lambda_{\rm 1, eff}(\Lambda_B)>\lambda_{\rm 1,min}(\Lambda_B)$, then the vacuum is metastable. 
\item
If $\lambda_{\rm 1, eff}(\Lambda_B)<\lambda_{\rm 1,min}(\Lambda_B)$, then the vacuum is unstable. 
\item
If $\lambda_2<0$, then the potential is unbounded from below along the $H, A$ and $H^\pm$ direction. 
\item
If $\lambda_3(\Lambda_{\rm I})<0$, the potential is unbounded from below along a direction in between $H^\pm$ and $h$.
\item
If $\lambda_L(\Lambda_{\rm I})<0$, the potential is unbounded from below along a direction in between $H$ and $h$.
\item
If $\lambda_S(\Lambda_{\rm I})<0$, the potential is unbounded from below along a direction in between $A$ and $h$.
\end{itemize}
In the above, $\Lambda_{\rm I}$ represents any energy scale for which $\lambda_{\rm 1, eff}$ is negative and the conditions for unboundedness of the potential follow from Eq.~(\ref{Scalarpot2}). At this point note the significant deviations we are making in the allowed parameter space compared to the usual vacuum stability conditions:  According to Eq.~(\ref{stabilitybound}), $\lambda_{3,L,S}$ can take slightly negative values. But at a scale where $\lambda_{\rm 1,eff}$ is negative, with the new conditions $\lambda_{3,L,S}$ have to be positive.

The tunneling probability ${\cal P}_0$ is computed by putting the minimum value of $\lambda_{\rm 1, eff}$ in Eq.~(\ref{action}) to minimize $S(\Lambda_B)$. In Fig.~\ref{fig:Tun}$\rm{ \left(a\right)}$, we have plotted  ${\cal P}_0$ as a function of  $M_t$. The right band corresponds to the tunneling probability for our benchmark point as in Table~\ref{table1}. For comparison, we plot ${\cal P}_0$ for SM as the left band in Fig.~\ref{fig:Tun}$\rm{(a)}$. 1$\sigma$ error bands in $\alpha_s$ and $M_h$ are also shown. The error due to $\alpha_s$ is more significant than the same due to $M_h$. As expected, for a given $M_t$, the presence of ID lowers tunneling probability. This is also reflected in Fig.~\ref{fig:Tun}$\rm {\left(b\right)}$, where we plot ${\cal P}_0$ as a function of  $\lambda_S(M_Z)$ for different choices of $\lambda_2(M_Z)$, assuming  $M_h=125.7$~GeV, $M_t=173.1$~GeV, and $\alpha_s=0.1184$. Here DM mass $M_A$ is also varied with $\lambda_S$ to get  $\Omega h^2=0.1198$. For a given $\lambda_S(M_Z)$, the higher the value of $\lambda_2(M_Z)$, the smaller ${\cal P}_0$ gets, leading to a more stable EW vacuum.

\section{Phase diagrams}  
\label{sec:phasediag}
The stability of EW vacuum depends on the value of parameters at low scale, chosen to be $M_Z$. In order to show the explicit dependence of EW stability on various parameters, it is customary to present phase diagrams in various parameter spaces.  

%%%%ContourSTU
 \begin{figure}[h!]
 \begin{center}
% \subfigure[]{
 \includegraphics[width=2.8in,height=2.8in, angle=0]{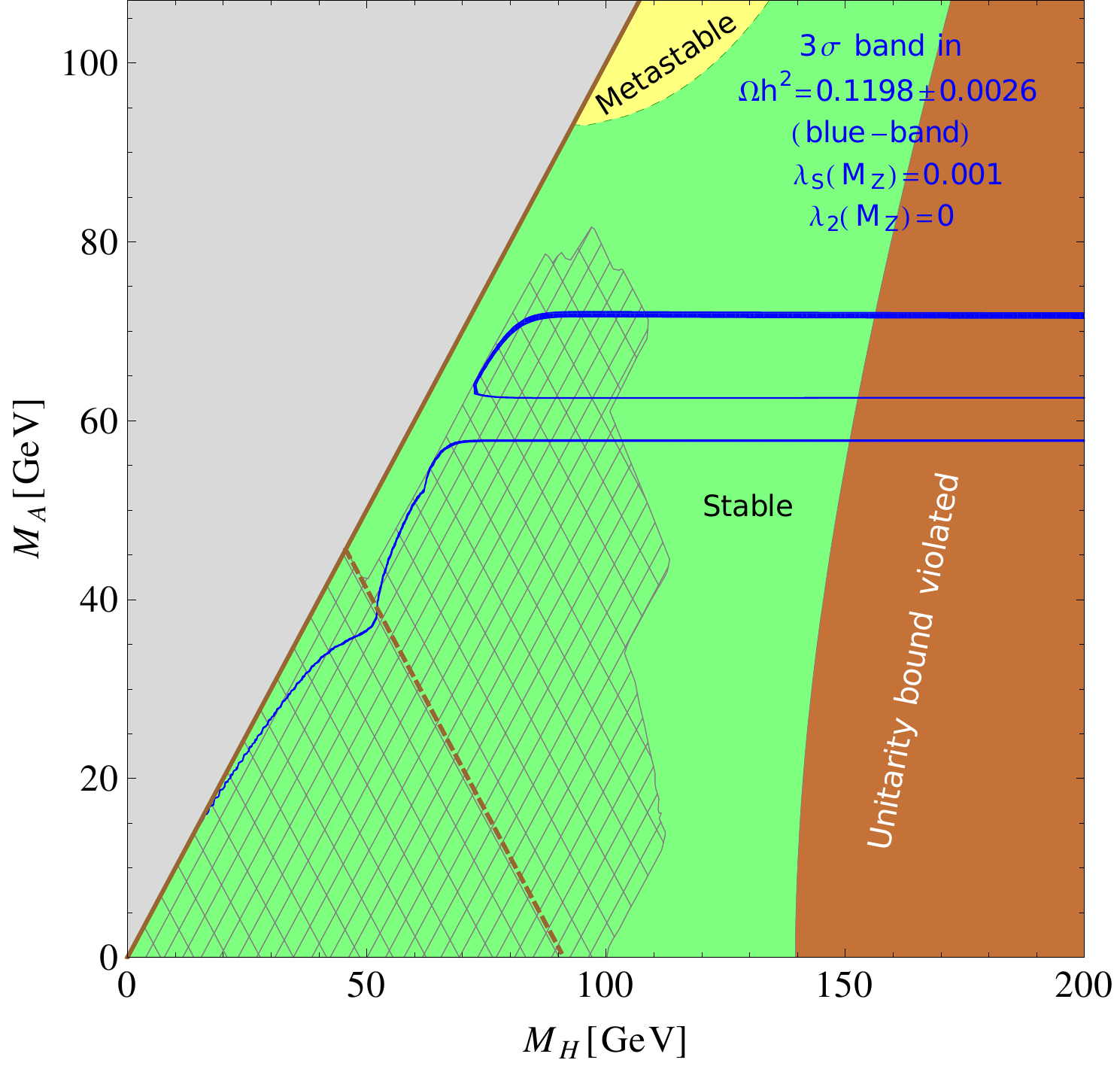}
% \hskip 15pt
% \subfigure[]{
% \includegraphics[width=2.8in,height=2.8in, angle=0]{metastableinertSTUplot70unitary}}
 \caption{\label{fig:STU1} \textit{\rm {Constraints in $M_{H}-M_{A}$ plane. The cross-hatched region  is excluded from LEP~{\rm \cite{Lundstrom:2008ai}}.  Choosing $M_{H^\pm}=120$~GeV and $\lambda_S(M_Z)=0.001$, relic density constraint is satisfied at 3$\sigma$ on the blue band. The green (yellow) region corresponds to EW vacuum stability (metastability). The solid brown line correspond to $M_H=M_A$. The grey area on the left to it is of no interest to us as we have chosen $M_H>M_A$. The dashed brown line shows the LEP\,I limit. On the brown region, unitarity constraints are violated before $\mpl$. } }}
 \end{center}
 \end{figure} 
%%%%ContourSTU
In Fig.~\ref{fig:STU1} we show the LEP constraints in the $M_{H}- M_A$ plane as in Ref.~\cite{Lundstrom:2008ai}. We update this plot identifying regions of EW stability and metastability. As we are considering a scenario where the ID model is valid till $\mpl$, there are further limits from unitarity.  The relic density constraint imposed by WMAP and Planck combined data is represented by the thin blue band. The choice of $\lambda_2(M_Z)$ does not have any impact on relic density calculations, but affects EW stability as expected. In this plot, for higher values of $\lambda_2(M_Z)$, the region corresponding to EW metastability will be smaller. The chosen parameters  satisfy the LUX direct detection bound.

%%%%ContourSTU
 \begin{figure}[h!]
 \begin{center}
% \subfigure[]{
% \includegraphics[width=2.8in,height=2.8in, angle=0]{MHvsMDMcontourplot}}
% \hskip 15pt
% \subfigure[]{
 \includegraphics[width=2.8in,height=2.8in, angle=0]{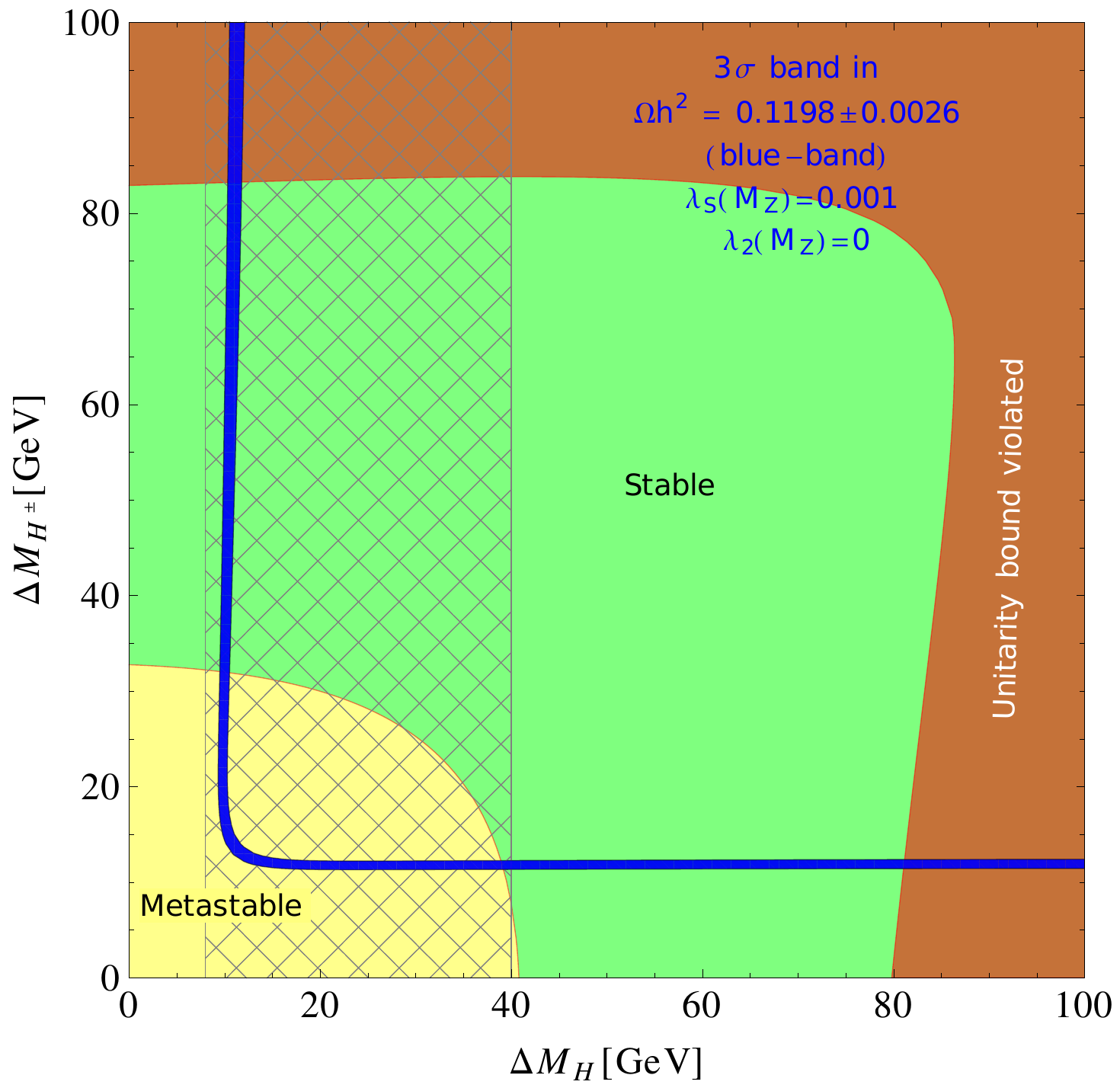}
 \caption{\label{fig:STU2} \textit{\rm {Phase diagram in $\Delta M_{H}-\Delta M_{H^\pm}$ plane for $M_A=70$~GeV. The green and yellow regions correspond to EW vacuum stability and metastability respectively. The cross-hatched band is excluded from LEP. The brown region suffers from unitarity violation before $\mpl$. The blue band reflects relic density constraint at 3$\sigma$.
  } }}
 \end{center}
 \end{figure} 
%%%%ContourSTU

As small splitting among $M_A$, $M_H$, and $M_{H^\pm}$ leads to some cancellations among diagrams contributing to DM annihilation, $\Delta M_{H^\pm}$ and $\Delta M_{H}$ are often used as free parameters in the ID model. In Fig.~\ref{fig:STU2}, we present constraints on this parameter space for $M_A=70$~GeV. As before, the brown region corresponds to unitarity violation before $\mpl$. For small $\Delta M_{H^\pm}$ and $\Delta M_{H}$, $\lambda_{3,4,5}$ are required to be small, which leads to little deviation from SM metastability. The metastable region is shown by the yellow patch, which shrinks for larger $\lambda_2$. The blue band reflects the relic density constraint for $\lambda_S(M_Z)=0.001$. For such small $\lambda_S(M_Z)$, the $h$-mediated $s$-channel diagram in $AA\ra WW$ or $AA\ra ZZ$ contributes very little. $H^+$- or $H$-mediated $t$- and $u$-channel diagrams are also less important than the quartic vertex driven diagram due to propagator suppression. This explains the ``L'' shape of the blue band. For higher values of $\lambda_S(M_Z)$, the shape of the band changes and ultimately leads to a closed contour. It appears that due to LEP constraints, EW vacuum metastability is almost ruled out. Although the LEP constraint permits $\Delta M_H<8$~GeV, allowing a narrow strip towards the left, the relic density constraints cannot be satisfied on this strip as it leads to an increased rate of DM coannihilation processes, leading to a dip in $\Omega h^2$. But as we will see later, if $M_t$ and $\alpha_s$ are allowed to deviate from their respective central values, for some region in this parameter space, it is possible to realize a metastable EW vacuum.

%%%%MDMvsLamSphase
 \begin{figure}[h!]
 \begin{center}
 \subfigure[]{
 \includegraphics[width=2.8in,height=2.8in, angle=0]{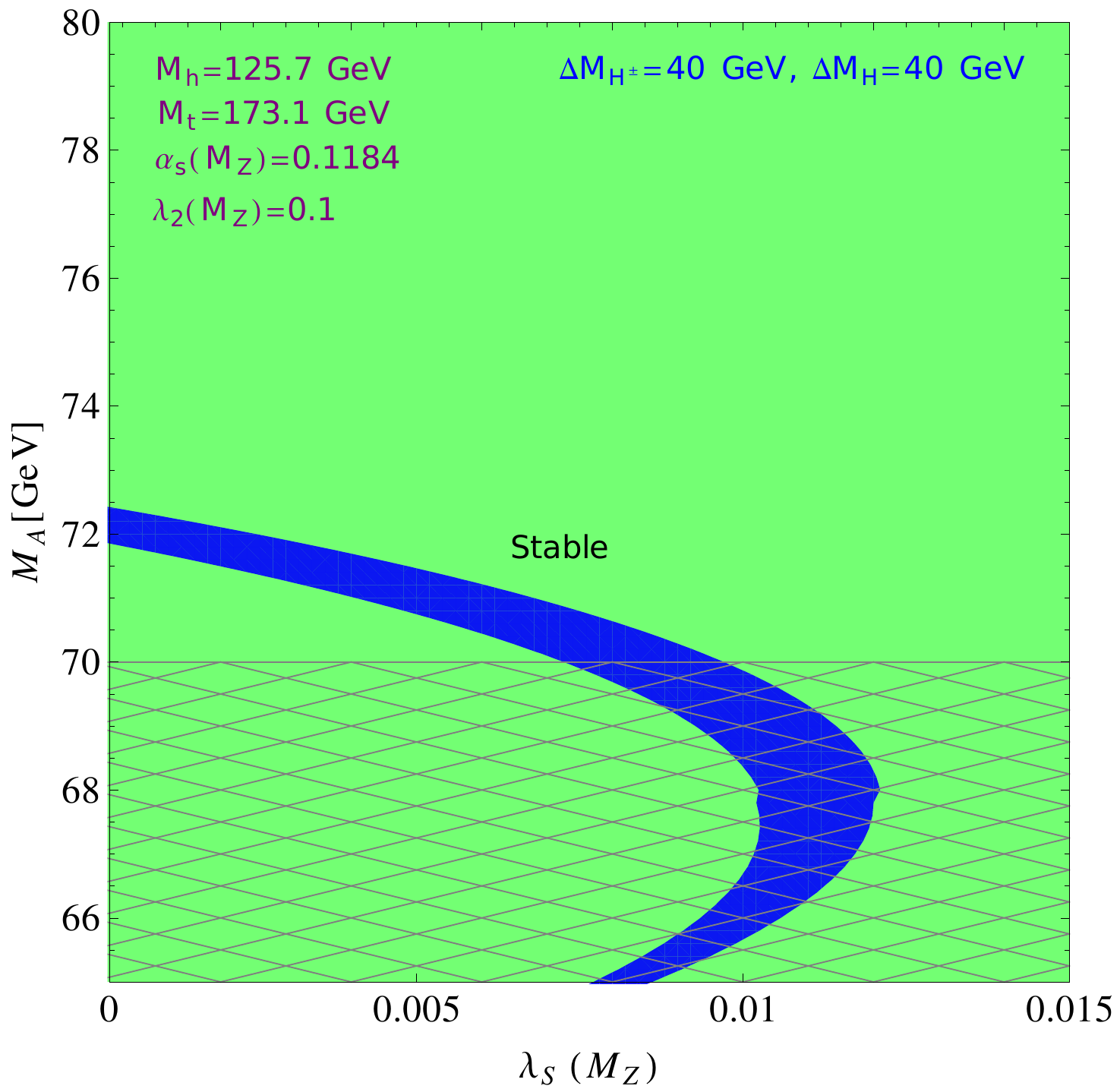}}
 \hskip 15pt
 \subfigure[]{
 \includegraphics[width=2.8in,height=2.8in, angle=0]{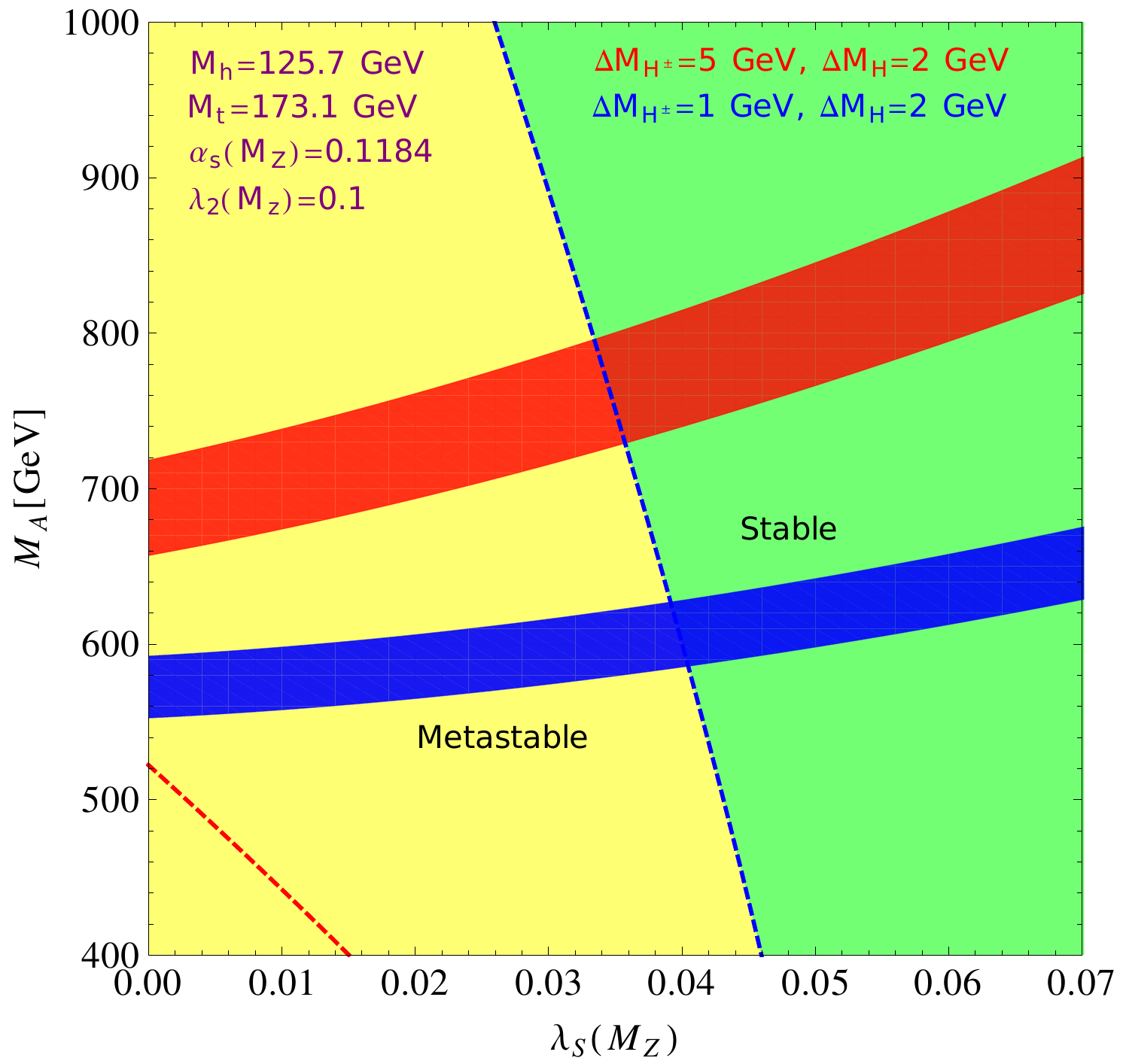}}
 \caption{\label{fig:MDM_LamS} \textit{\rm {Phase diagram in $\lambda_S(M_Z)-M_A$ plane for $\lambda_2(M_Z)=0.1$. Panel (a) stands for `low' DM mass. The blue band  corresponds the 3$\sigma$ variation in $\Omega h^2$ when $\Delta M_{H^\pm}=40$~GeV and $\Delta M_H=40$~GeV. LEP direct search constraints are represented by the cross-hatched band at the bottom. Entire green region imply EW vacuum stability. Panel (b) stands for `high' DM masses.  The relic density band (blue) now correspond to  $\Delta M_{H^\pm}=1$~GeV and $\Delta M_H=2$~GeV. The corresponding stable and metastable phases for EW vacuum are represented by green and yellow patches respectively.   The relic density band (red) corresponds to  $\Delta M_{H^\pm}=5$~GeV and $\Delta M_H=2$~GeV. For this, the boundary separating the EW phases is denoted by the red dashed line.} }}
 \end{center}
 \end{figure}
%%%%MDMvsLamSphase

To delineate the role of $M_H$ in EW vacuum stability, in Figs.~\ref{fig:STU1} and \ref{fig:STU2}, $\lambda_2(M_Z)$ was chosen to be small. Now to demonstrate the effect of $\lambda_S$, we will now present in Fig.~\ref{fig:MDM_LamS} phase diagrams in the $\lambda_S(M_Z) - M_A$ plane. Panel (a) deals with low DM masses. For $\Delta M_{H^\pm}=40$~GeV and $\Delta M_H=40$~GeV, part of the allowed relic density band (blue) is allowed from LEP constraints (cross-hatched band). The entire parameter space corresponds to EW vacuum stability. Choosing small  $\Delta M_{H^\pm}$ and $\Delta M_H$, which imply small values of $\lambda_{3,4,5}$, can lead to metastability. But those regions are excluded by LEP. Again, metastability can creep in if $M_t$ and $\alpha_s$ are allowed to deviate from  their central values. 

In Fig.~\ref{fig:MDM_LamS}(b), we study the same parameter space for high DM masses. As mentioned before, to obtain the correct relic density, smaller mass splitting among various ID scalars needs to be chosen. For $\Delta M_{H^\pm}=1$~GeV and $\Delta M_H=2$~GeV, the 3$\sigma$ relic density constraint is shown as the blue band. The blue dashed line demarcates the boundary between stable (green) and metastable (yellow) phases of EW vacuum. The choice of small values of $\Delta M_{H^\pm}$ and $\Delta M_H$, in turn, leads to a large region pertaining to EW metastability. 

To illustrate the sensitivity to the mass splitting, in Fig.~\ref{fig:MDM_LamS}(b), we present another relic density  band (red) when $\Delta M_{H^\pm}=5$~GeV and $\Delta M_H=2$~GeV. The corresponding boundary between the phases is denoted by the red dashed line. The region on the right implies EW stability (the green and yellow regions do not apply to this case). As for high DM masses, EW metastability can be attained for a sizable amount of the parameter space; $\lambda_2(M_Z)$ need not be chosen to be very small to maximize the metastable region for the sake of demonstration. 

%%%%mtmhphase
 \begin{figure}[h!]
 \begin{center}
 \subfigure[]{
 \includegraphics[width=2.8in,height=2.8in, angle=0]{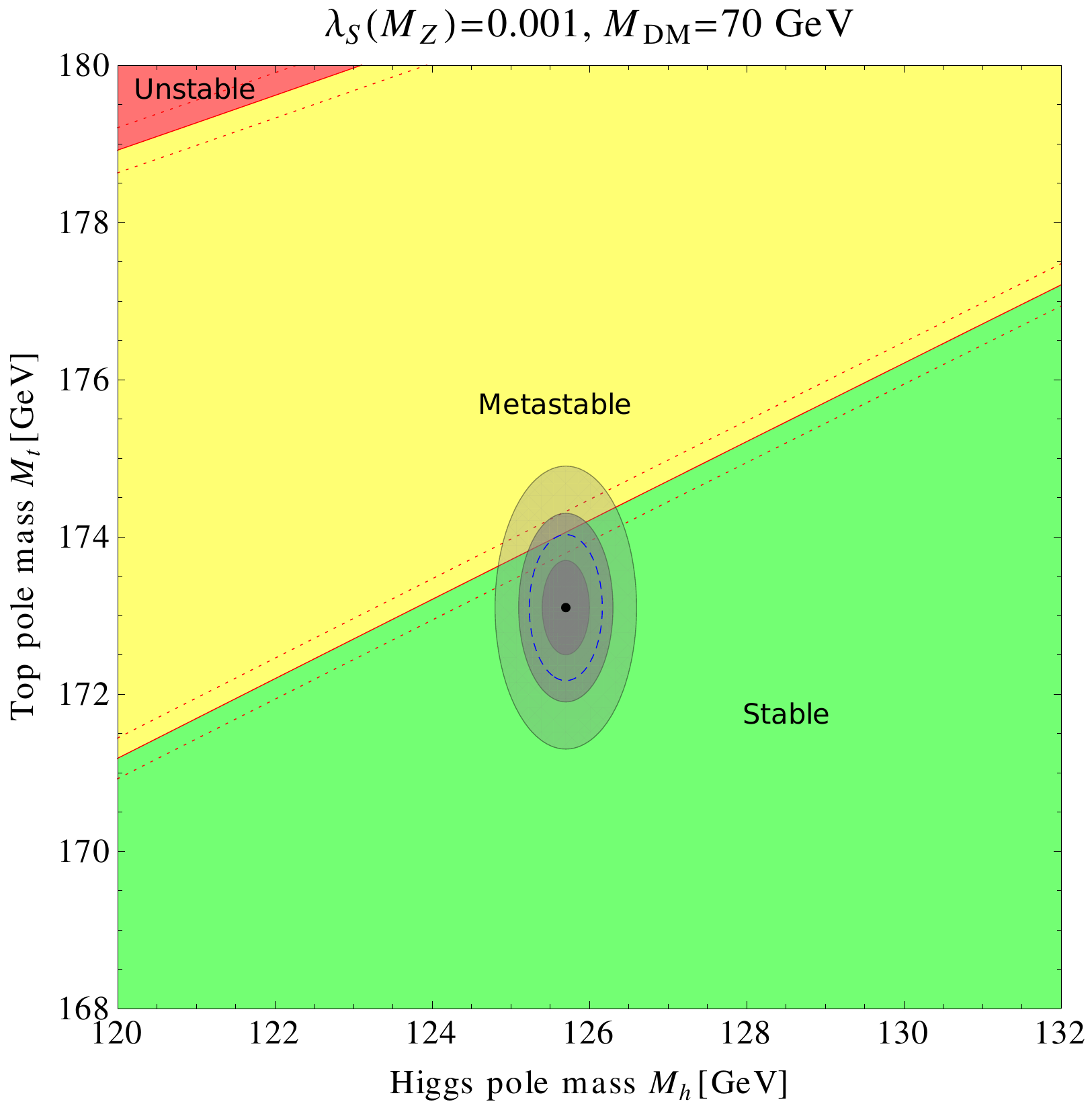}}
 \hskip 15pt
 \subfigure[]{
 \includegraphics[width=2.8in,height=2.8in, angle=0]{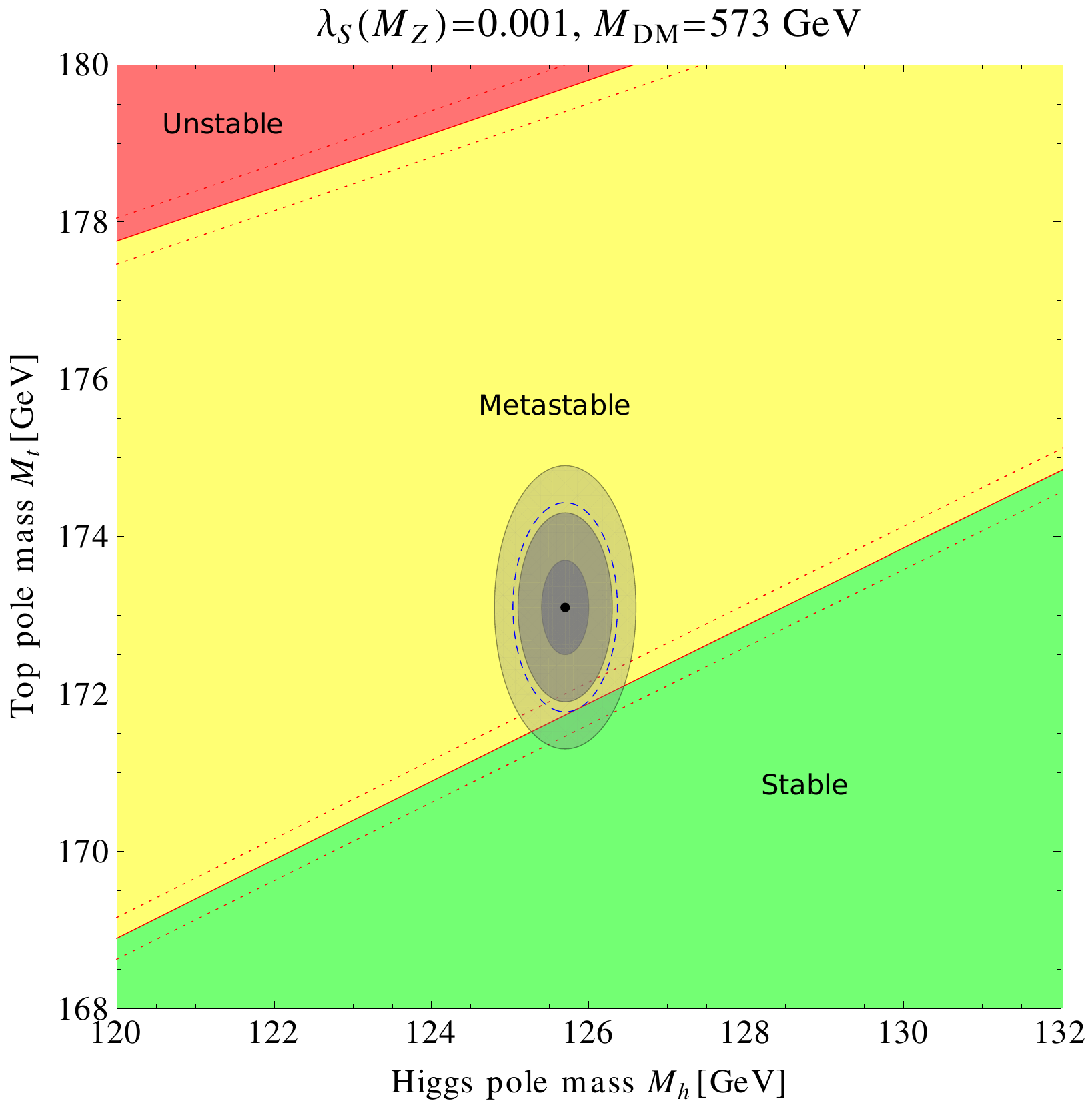}}
 \caption{\label{fig:Mt_Mh} \textit{\rm {Phase diagrams in $M_h - M_t$ plane. Panels (a) and (b) stand for `low' and `high' DM masses respectively. Regions of 
absolute stability~(green), metastability~(yellow), instability~(red) of the EW vacuum are also marked. The grey zones represent error ellipses at  $1$, $2$ and  $3\sigma$. The three boundary lines (dotted, solid and dotted red) correspond to $\alpha_s(M_Z)=0.1184 \pm 0.0007$. Details of benchmark points are available in the text. } }}
 \end{center}
 \end{figure}
%%%%mtmhphase

The fact that for SM the EW vacuum stability is ruled out at $\sim 3\sigma$, is demonstrated by a phase space diagram in the $M_t-M_h$ plane~\cite{Degrassi:2012ry, Buttazzo:2013uya}. In Ref.~\cite{Khan:2014kba}, similar diagrams were presented for a singlet scalar extended SM. To demonstrate the impact of ID scalars to uplift the EW vacuum metastability, we present phase diagrams in the $M_t-M_h$ plane for two sets of benchmark points in Fig.~\ref{fig:Mt_Mh}. Panel (a) is drawn for $M_A = 70$~GeV, $\Delta M_{H^\pm}=11.8$~GeV, $\Delta M_H=45$~GeV, $\lambda_S(M_Z)=0.001$, and $\lambda_2 (M_Z) = 0.1$. For panel (b) the set of parameters in Table~\ref{table1} is being used. 
Both sets of parameters are chosen so that they respect the WMAP and Planck combined results on DM relic density and the direct detection bounds from XENON\,100 and LUX. As in Ref.~\cite{Khan:2014kba}, the line demarcating the boundary between stable and metastable phases of EW vacuum is obtained by demanding that the two vacua be at the same depth, implying $\lambda_1(\Lambda_B)=\beta_{\lambda_1}(\Lambda_B)=0$. The line separating the metastable phase from the unstable one is drawn using the conditions $\beta_{\lambda_1}(\Lambda_B)=0$ and $\lambda_1(\Lambda_B)=\lambda_{1, \rm min}(\Lambda_B)$, as in Eq.~(\ref{lammin}). The variations due to uncertainty in the measurement of $\alpha_s$ are marked as dotted red lines. In each panel, the dot representing central values for $M_h$ and $M_t$ is encircled by 1$\sigma$, 2$\sigma$, and 3$\sigma$ ellipses representing errors in their measurements.  According to Fig.~\ref{fig:Mt_Mh}(a), EW vacuum stability is allowed at 1.5$\sigma$, whereas in Fig.~\ref{fig:Mt_Mh}(b), it is excluded at 2.1$\sigma$, indicated by blue-dashed ellipses.

%%%%Mt-alphaS phase diagram
  \begin{figure}[h!]
 \begin{center}
 \subfigure[]{
 \includegraphics[width=2.8in,height=2.8in, angle=0]{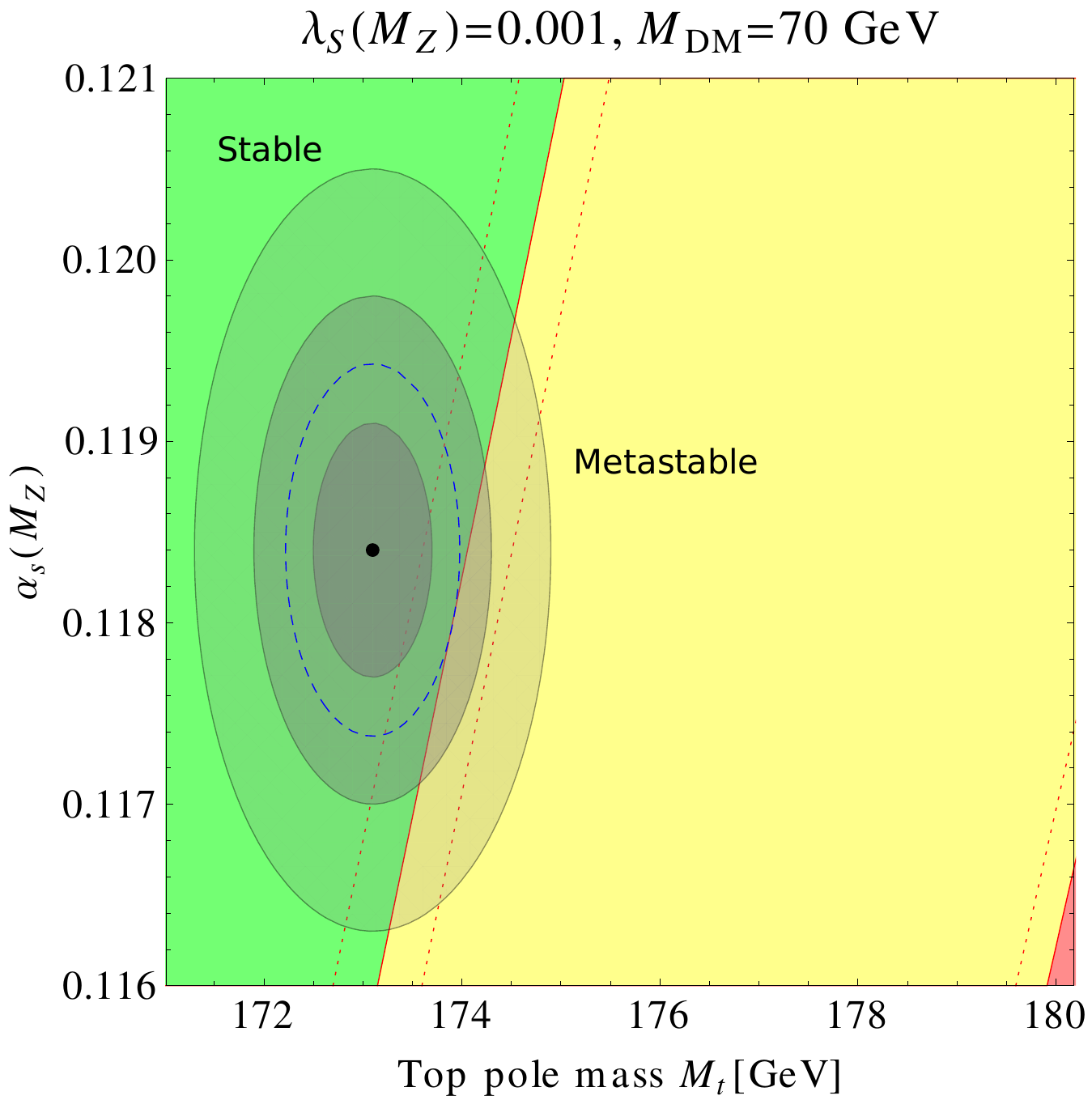}}
 \hskip 15pt
 \subfigure[]{
 \includegraphics[width=2.8in,height=2.8in, angle=0]{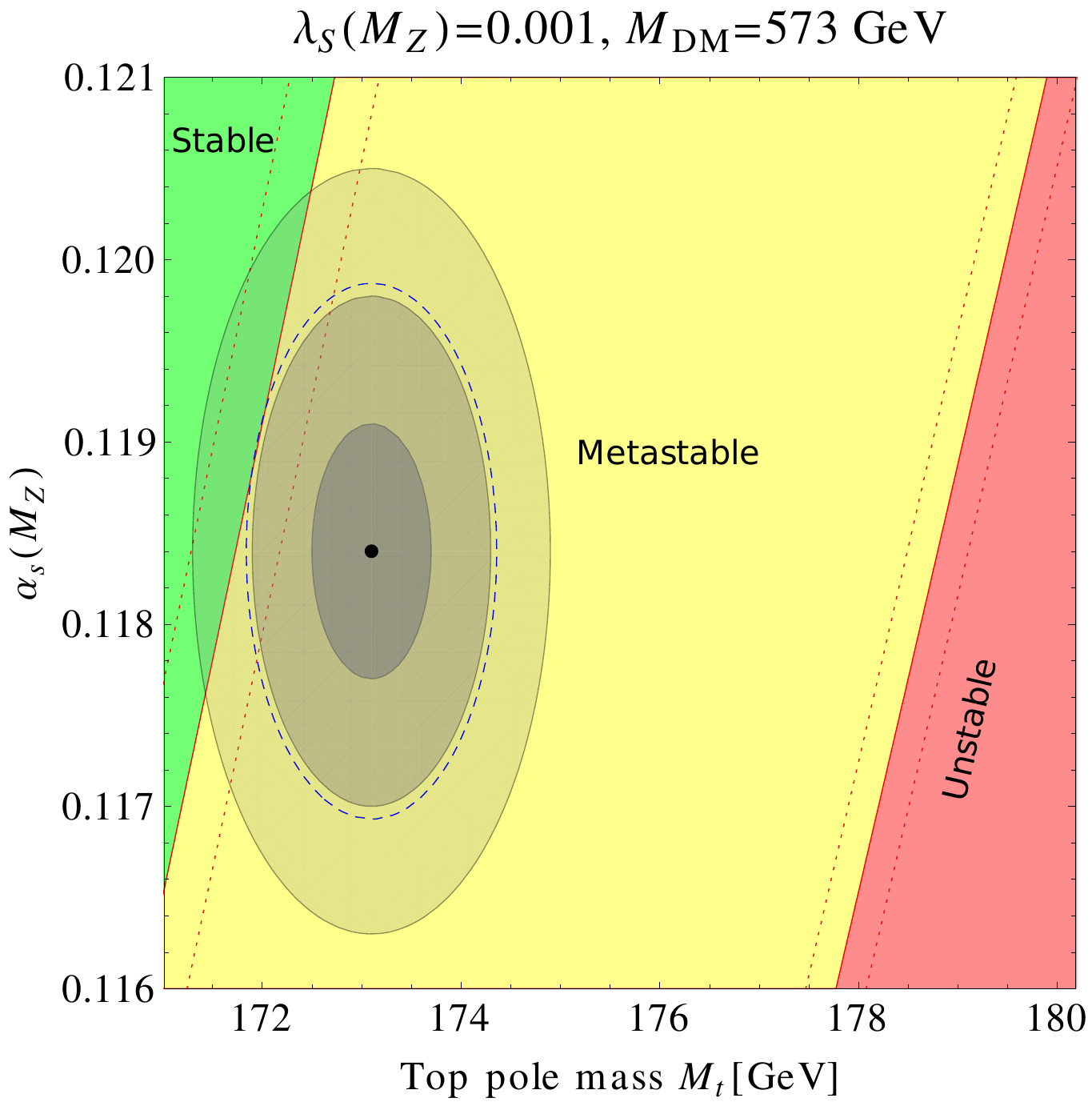}}
 \caption{\label{fig:Alpha_Mt} \textit{\rm {Phase diagrams in $M_t-\alpha_s(M_Z)$ plane  for the same sets of benchmark points as in Fig.~\ref{fig:Mt_Mh}. Notations used are also the same as in Fig.~\ref{fig:Mt_Mh}. } }}
 \end{center}
 \end{figure}
%%%%Mt-alphaS phase diagram 
As in the literature SM EW phase diagrams are also presented in the $\alpha_s(M_Z)-M_t$ plane~\cite{Bezrukov:2012sa, EliasMiro:2011aa}, we do the same in the ID model as well. In Fig.~\ref{fig:Alpha_Mt}, we use the same sets of benchmark parameters as in Fig.~\ref{fig:Mt_Mh}. As a consistency check, one can note that the EW vacuum is allowed or ruled out at the same confidence levels.

%%%%confidence
 \begin{figure}[h!]
 \begin{center}{
 \includegraphics[width=2.8in,height=2.8in, angle=0]{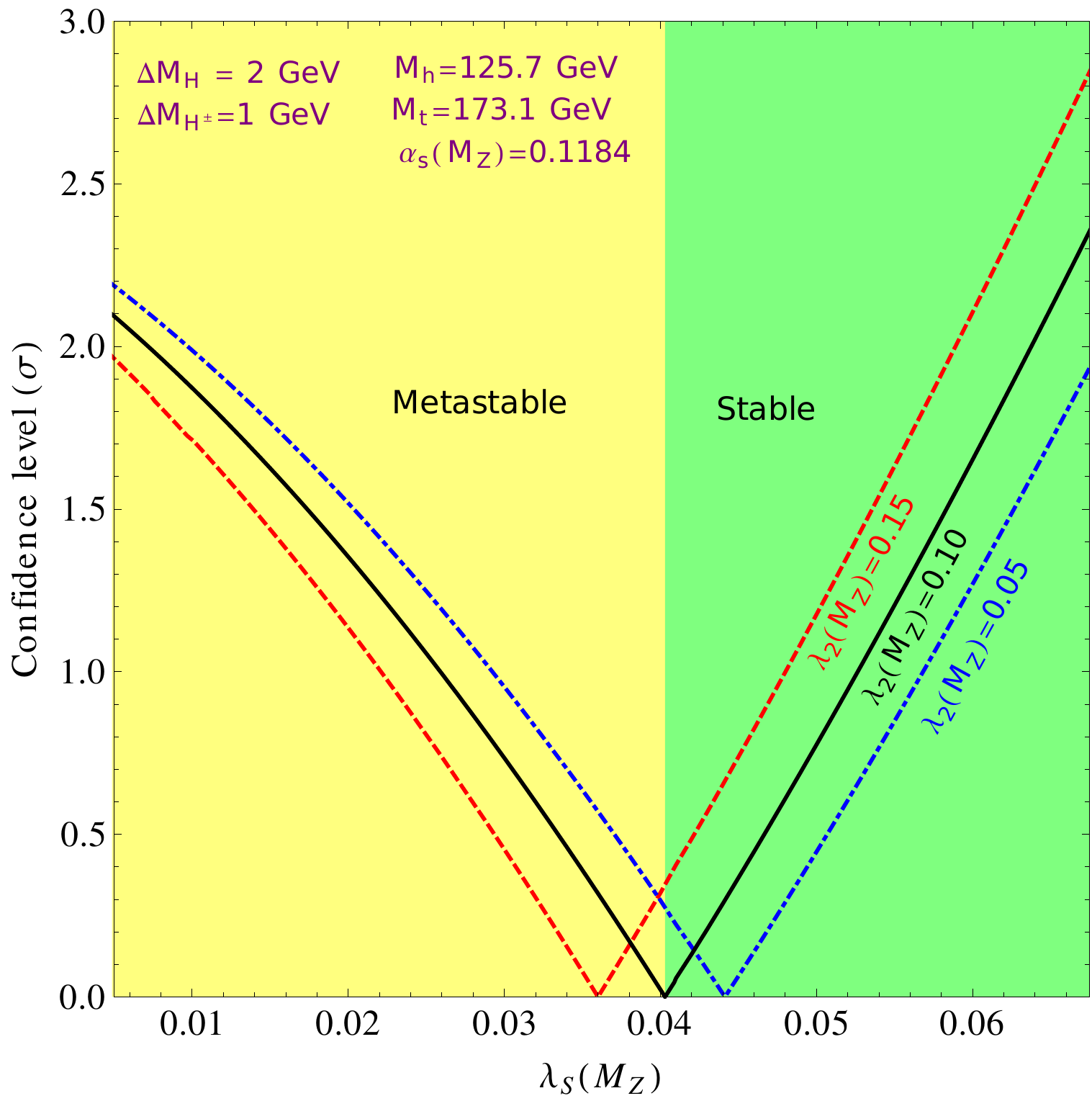}}
 \caption{\label{fig:confidence} \textit{\rm {Dependence of confidence level  at which EW vacuum stability is excluded (one-sided) or allowed on $\lambda_S(M_Z)$ and $\lambda_2(M_Z)$. Regions of absolute stability (green) and metastability (yellow) of EW vacuum are shown for $\lambda_2(M_Z)=0.1$. The positive slope of the line corresponds to the stable electroweak vacuum and negative slope corresponds to the metastability.     
 } }}
 \end{center}
 \end{figure}
%%%%confidence
To study the impact of nonzero ID couplings, however, it is instructive to study the change in the confidence level ($\sigma$) at which EW stability is modified with respect to these couplings. As in Ref.~\cite{Khan:2014kba}, we plot in Fig.~\ref{fig:confidence} ~$\sigma$ against $\lambda_S(M_Z)$ for different values of $\lambda_2(M_Z)$. We vary $M_A$ along with $\lambda_S(M_Z)$ to keep DM relic density fixed at $\Omega h^2=0.1198$ throughout the plot. Note that changing $\lambda_2(M_Z)$ does not alter $\Omega h^2$. The masses of other ID particles  are determined using $\Delta M_{H^\pm}=1$~GeV and $\Delta M_H=2$~GeV. The parameter space considered does not yield too large DM-nucleon cross section, inconsistent with XENON\,100 and LUX DM direct detection null results. For a specific value of $\lambda_2(M_Z)=0.1$, with the increase of $\lambda_S(M_Z)$, the confidence level at which EW is metastable (yellow region) gets reduced and becomes zero at $\lambda_S(M_Z)\simeq 0.04$. After this, EW vacuum enters in the stable phase (green). With further increases in $\lambda_S(M_Z)$, the confidence level at which EW is stable keeps increasing. To illustrate the role of $\lambda_2(M_Z)$,  we use two other values in the same plot. The value of $\lambda_S(M_Z)$ at which the EW vacuum enters in the stable phase increases with decreases in $\lambda_2(M_Z)$, as expected. The yellow and green marked regions are not applicable when $\lambda_2(M_Z)=0.05, 0.15$.

\section{Veltman's conditions}
\label{sec:veltman}

As we have extended the validity of the ID model till $\mpl$, it is interesting to explore whether Veltman's condition (VC) can be satisfied in this model at any scale on or before $\mpl$. It is particularly interesting as similar studies for SM have been carried out in Ref.~\cite{Degrassi:2012ry}. It has been shown that if one imposes VC in the SM at $\mpl$, then the top mass measurement $M_t=173.1\pm 0.6$~GeV implies $M_h\approx 135\pm2.5$~GeV, which is excluded at more than 3$\sigma$.

Veltman's condition implies that the quadratic divergences in the radiative corrections to the Higgs mass can be handled if the coefficient multiplying the divergence somehow vanishes~\cite{Veltman:1980mj, Hamada:2012bp}. VC includes the contributions from the infrared degrees of freedom of the theory and does not carry any special information about the ultraviolet divergences. In SM, it suggests the combination 
\beq
6 \lambda_1+\frac{9}{4} g_2^2+\frac{3}{4} g_1^2 -12 y_t^2 = 0\nn\, .
\eeq
Due to the large negative contribution from the term containing the top Yukawa coupling, it is not possible to satisfy VC till $\mpl$ given the experimental measurements of $M_t$ and $M_h$ within the context of SM.

In the ID model, as we add more scalars a possibility opens up to satisfy VC, as their contributions can offset the large negative contribution from the top quark. 

The above VC for the SM associated with $\mu_1$ is promoted in the ID model to~\cite{Barbieri:2006dq, Chakraborty:2014oma}
\beq
6 \lambda_1+ 2\lambda_3+\lambda_4+\frac{9}{4} g_2^2+\frac{3}{4} g_1^2 -12 y_t^2  = 0.
\eeq
If $2\lambda_3+\lambda_4$ is positive, then we have checked with our RG improved coupling constants, and it is possible to satisfy the above VC at a scale before $\mpl$. 

However, in the ID model $\mu_2$ also receives quadratically divergent radiative corrections. The corresponding VC reads as
\beq
6 \lambda_2+ 2\lambda_3+\lambda_4+\frac{9}{4} g_2^2+\frac{3}{4} g_1^2 = 0\, .
\eeq
Note that it lacks the Yukawa contribution as the unbroken $\Z_2$ forbids fermionic interactions of the inert doublet $\Phi_2$. As $2\lambda_3+\lambda_4$ is already positive, this VC can be satisfied if $\lambda_2$ is negative. But a negative $\lambda_2$ renders the potential unbounded from below as evident from our earlier discussions. Note that amongst our RG improved coupling constants $\lambda_1$ can be driven to negative values at high scales. But this makes the required cancellations for VCs even worse. Hence, it is not possible to satisfy Veltman's conditions in a scenario where only the ID model reigns the entire energy regime up to the $\mpl$.

\section{Summary and Conclusion}
\label{sec:conclusion}
If the standard model is valid up to the Planck scale, the present measurements on the masses of the top quark and Higgs indicate the presence of a deeper minimum of the scalar potential at a very high energy scale, threatening the stability of the present electroweak vacuum.  State of the art NNLO calculations performed to evaluate the probability that the present EW vacuum will tunnel into the deeper vacuum lying close to $\mpl$ suggest that the present EW vacuum is metastable at $\sim 3\sigma$. The lack of stability might be the artifact of incompleteness of the SM. 

Although the LHC has yet to find any signal suggesting existence of any new physics beyond the standard model of particle physics, other experimental evidence points towards the existence of dark matter, which so far could have escaped detection in colliders and DM direct detection experiments. Hence, it is important to look into the problem of EW stability in a scenario which addresses the issue of DM as well. In particular, we extend SM by adding an inert scalar doublet, offering a viable DM candidate and assume that this model is valid up to $\mpl$. 

In this paper, our intentions are twofold. First, in such a scenario we have consolidated the bounds imposed on the ID model. As we are demanding validity of the model up to $\mpl$, the RG evolution of the couplings can disturb the unitarity of the S-matrix governing various scattering processes, which in turn imposes stringent limits on the parameter space at the EW scale. In this light, we present a consolidated discussion updating the existing bounds on the ID model.

The other and the main goal of this paper is to check the stability of EW vacuum in the ID model.   If the ID DM happens to be the only DM particle, which saturates the observed DM relic density, can the ID model modify the stability of the EW vacuum? Note that rather than considering new physics effects close to the Planck scale, here new physics is added at the EW scale only.
 It is well known that addition of a scalar can improve stability of the EW vacuum. But if we are to solve both the DM and EW vacuum stability problems in the context of the ID model, it is important to study the parameter space which allows us to do so. As ID introduces a few new parameters and fields, the study of the parameter space is quite involved when we consider radiatively improved scalar potentials containing SM NNLO corrections and two-loop ID contributions. Inclusion of these NNLO corrections is mandatory to reproduce the correct confidence level at which EW vacuum is metastable in the SM. In case of ID, if one works with one-loop ID contributions in the RGEs instead of two-loop ones, the changes in the plots presented in the paper are limited to $5$\% only. However, as a small change in the action $S$ is amplified exponentially in the vacuum decay lifetime, two-loop effects have also been taken care of. For the benchmark point used in Table~\ref{table1}, ${\cal P}_0$ using one-loop ID RGEs turns out to be $5.8 \times 10^{-1271}$, whereas for two-loop ID RGEs,  ${\cal P}_0$ changes to $1.4 \times 10^{-1310}$. We neglect other loop corrections to bounce action, as a $5$\%--$10$\% correction can lead to an appreciable change in the lifetime, but as demonstrated above with the example of inclusion of two-loop ID corrections, the phase diagrams essentially remain unchanged (see Ref.~\cite{Branchina:2014rva} for a related example). We note in passing that unlike Ref.~\cite{Goudelis:2013uca}, our analysis is valid up to $\mpl$. Requiring the potential bounded from below at all scales below $\mpl$ puts severe  constraints on the model.
 
As far as the allowed parameter space is concerned, we focus on metastability of EW vacuum which is realized when $\lambda_1$ is negative.  In this context, our analysis is much different from earlier analyses in the literature, for example, as in Ref.~\cite{Goudelis:2013uca}. When $\lambda_1$ is negative, the vacuum stability conditions (\ref{stabilitybound}) are no longer useful and the parameter space analysis in Ref.~\cite{Goudelis:2013uca} is not valid. For example, when $\lambda_1$ is positive, $\lambda_{3,L,S}$ can assume small negative values as positivity of the $\lambda_1 h^4$ term ensures the potential bounded from below. However as we point out, one can no longer afford a negative $\lambda_{3,L,S}$ at a scale where $\lambda_1$ is negative. If we demand in addition that the ID model saturates DM relic density, then 
even when $\lambda_1$ is positive, $\lambda_{3}$ cannot be negative. But as mentioned earlier, for  negative $\lambda_{1}$, although $\lambda_{3}$ has to be positive at that scale, its value at $M_Z$ can still be negative.  Thus, one of the main motives for this work is to find out the  allowed parameter space in the ID DM model in case the EW vacuum is metastable. 
 
 We see that for DM masses of 70~GeV, the allowed parameter space corresponds to absolute stability  unless we allow some deviation of $M_t$ and $\alpha_s$ from their measured central values. For higher DM masses more than 500~GeV, it is possible to realize a metastable EW vacuum for a large parameter space. This vacuum will have a longer lifespan than the SM one as the addition of scalars improves the stability of the EW vacuum.

\vskip 20pt
%\section{Acknowledgements}
\noindent{\bf Acknowledgements:}\\
The work of N.K. is supported by a fellowship from UGC. S.R. is indebted to Sourov Roy and Dilip Ghosh for useful discussions.

\appendix

\section{One-loop beta functions}
\label{App:BetaFunctions}
The beta functions of the quartic coupling parameters for the ID model are defined as 
\beq
\beta_{\lambda_{i}}=16\pi^2 \frac{\partial\lambda_{i}}{\partial \ln \mu}\, . 
\eeq 
The expressions at one loop are given by~\cite{Ferreira:2009jb,Goudelis:2013uca}
\bea
\beta_{\lambda_1} &=& 24 \lambda_1^2 + 2 \lambda_3^2 + 2 \lambda_3 		\lambda_4 +  \lambda_4^2 +  \lambda_5^2 \nn \\
	&&+\frac{3}{8} \left( 3 g_2^4 + g_1^4 + 2 g_2^2 g_1^2 \right) - 3 \lambda_1 \left( 3 g_2^2 + g_1^2 \right) \nn \label{betal_1}\\	
	&&+ 4 \lambda_1 \left( y_\tau^2 + 3 y_b^2 + 3 y_t^2 \right) - 2 \left( y_\tau^4 + 3 y_b^4 + 3 y_t^4 \right), \\
\beta_{\lambda_2} &=& 24 \lambda_2^2 + 2 \lambda_3^2 + 2 \lambda_3 \lambda_4 +  \lambda_4^2 +  \lambda_5^2 \nn \\
	&&+\frac{3}{8} \left( 3 g_2^4 + g_1^4 + 2 g_2^2 g_1^2 \right) - 3 \lambda_2 \left( 3 g^2 + g_1^2 \right) , \\	
\beta_{\lambda_3} &=& 4\left( \lambda_1 + \lambda_2 \right) \left( 3 \lambda_3 +  \lambda_4 \right) + 4 \lambda_3^2 + 2 \lambda_4^2 + 2 \lambda_5^2 \nn \\
	&&+	\frac{3}{4} \left( 3 g_2^4 + g_1^4 - 2 g_2^2 g_1^2 \right) - 3 \lambda_3 \left( 3 g_2^2 + g_1^2 \right) \nn \\
	&&+2 \lambda_3 \left( y_\tau^2 + 3 y_t^2 + 3 y_b^2 \right) , \\	
\beta_{\lambda_4} &=& 4 \lambda_4 \left( \lambda_1 + \lambda_2 + 2 \lambda_3 +  \lambda_4 \right) + 8 \lambda_5^2 \nn\\
	&&+ 3 g_2^2 g_1^2 - 3 {\lambda_4} \left( 3 g_2^2 + g_1^2 \right) \nn \\
	&&+ 2 \lambda_4 \left( y_\tau^2 + 3 y_t^2 + 3 y_b^2 \right) , \\   
\beta_{\lambda_5} &=& 4 \lambda_5 \left( \lambda_1 + \lambda_2 + 2 \lambda_3 + 3 \lambda_4 \right) \nn\\
	&& - 3 \lambda_5 \left( 3 g_2^2 + g_1^2 \right)  \nn \\
	&& +2 \lambda_5 \left( y_\tau^2 + 3 y_t^2 + 3 y_b^2 \right).
\eea
Let us note the $y_t$ dependence of these expressions. While $\beta_{\lambda_1}$ is dominated by the $y_t^4$ term,  $\beta_{\lambda_2}$ does not depend on $y_t$. The $y_t$ dependence of other $\beta_{\lambda_i}$s are softened by the corresponding $\lambda_i$ multiplying the $y_t^2$ terms. Two-loop RGEs used in this work have been generated using {\tt SARAH}~\cite{Staub:2013tta}.

\end{document}